\PassOptionsToPackage{table}{xcolor}
\documentclass[twocolumn,prd,preprintnumbers,numbers,sort&compress,nofootinbib,showpacs,colorlinks,citecolor=blue,amsmath,amssymb,aps,superscriptaddress]{revtex4-2}

\usepackage{graphicx}
\usepackage{dcolumn}
\usepackage{bm}
\usepackage{xcolor}
\usepackage{lipsum} 
\usepackage{float} 

\newcommand{\be}{\begin{equation}}
\newcommand{\ee}{\end{equation}}
\newcommand{\bw}{\begin{widetext}}
\newcommand{\ew}{\end{widetext}}
\newcommand{\ba}{\begin{aligned}}
\newcommand{\ea}{\end{aligned}}
\newcommand{\bes}{\begin{equation*}}
\newcommand{\ees}{\end{equation*}}
\newcommand{\bea}{\begin{eqnarray}}
\newcommand{\eea}{\end{eqnarray}}
\newcommand{\dd}{\text{d}}

\newcommand{\hmC}{\hat{\mathcal{C}}}

\newcommand{\fR}{\mathfrak{R}}
\newcommand{\hfR}{\hat{\mathfrak{R}}}
\newcommand{\tFD}{\mathcal{T}_\mathrm{FD}}

\newcommand{\beq}{\begin{equation}}
\newcommand{\eeq}{\end{equation}}
\newcommand{\exclude}[1]{}
\newcommand{\jcap}{JCAP}
\newcommand{\apjl}{ApJL~}
\newcommand{\aap}{Astronomy \& Astrophysics}

\newcommand{\mnras}{MNRAS}
\newcommand{\apjs}{ApJ\ Supp.}

\newcommand{\araa}{ARA \& A}

\newcommand{\physrep}{Physics Reports}

\definecolor{Orange}{rgb}{1.0,0.5,0.15}
\definecolor{Blue}{rgb}{0,0.08,0.65}
\definecolor{Red}{rgb}{0.65,0.08,0.05}
\definecolor{Green}{rgb}{0.15,0.45,0.25}
\definecolor{Pink}{rgb}{1.0,0.05,0.5}
\definecolor{bubbles}{rgb}{0.91, 1.0, 1.0}
\definecolor{aquamarine}{rgb}{0.5, 1.0, 0.83}
\definecolor{bubblegum}{rgb}{0.99, 0.76, 0.8}
\definecolor{bluebell}{rgb}{0.74, 0.74, 0.92}
\definecolor{dollarbill}{rgb}{0.72, 0.93, 0.6}
\definecolor{cred}{RGB}{238,28,37}

\newcommand{\pos}[1]{ {\color{cred}#1}}

\newcommand{\ubc}{Department of Physics \& Astronomy, University of British Columbia, 6224 Agricultural Road, Vancouver, BC V6T 1Z1, Canada}

\newcommand{\bochum}{Astronomisches Institut, Ruhr-Universität Bochum, Universitätsstr. 150, 44801, Bochum, Germany}

\newcommand{\cea}{Universit\'e Paris-Saclay, IPHT, DRF-INP, UMR 3681, CEA, Orme des Merisiers Bat 774, 91191 Gif-sur-Yvette, France}

\newcommand{\cgpqi}{Center for Gravitational Physics and Quantum Information, Yukawa Institute for Theoretical Physics, Kyoto University, Kyoto 606-8502, Japan}

\newcommand{\aas}{American Astronomical Society, 1667 K St NW, Washington, DC 20006}

\newcounter{FFcounter}

\newcounter{LVWcounter}

\newcounter{SGcounter}

\newcounter{RDcounter}

\usepackage{hyperref}

\begin{document}

\preprint{APS/123-QED}

\title{Mitigating Nonlinear Systematics in Weak Lensing Surveys: The Bernardeau-Nishimichi-Taruya Approach}

\author{Shiming Gu}
 \email{gsm@phas.ubc.ca}
\affiliation{\ubc}
\affiliation{\bochum}

\author{Ludovic van Waerbeke}%
 \email{waerbeke@phas.ubc.ca}
\affiliation{\ubc}

\author{Francis Bernardeau}
      \affiliation{\cea}
      \affiliation{\cgpqi}

\author{Roohi Dalal}
\affiliation{\ubc}
\affiliation{\aas}

\date{\today}

\begin{abstract}
Weak lensing surveys, along with most other late-Universe probes, have consistently measured a lower amplitude of the matter fluctuation spectrum, denoted by the parameter $S_8$, compared to predictions from early-Universe measurements in cosmic microwave background data. 
Improper modeling of nonlinear scales may partially explain these discrepancies in lensing surveys.
This study investigates whether the conventional approach to addressing small scale biases remains optimal for Stage-IV lensing surveys. We demonstrate that conventional weak lensing estimators are affected by scale leakage from theoretical biases at nonlinear scales, which influence all observed scales. Using the BNT transform, we propose an $\ell$-cut methodology that effectively controls this leakage. The Bernardeau-Nishimichi-Taruya (BNT) transform reorganizes weak lensing data in $\ell$ space, aligning it with $k$ space, thereby reducing the mixing of nonlinear scales and providing a more accurate interpretation of the data.
We evaluate the BNT approach by comparing \textsc{HMcode}, \textsc{Halofit}, \textit{Baryon Correction Model} and \textsc{AxionHMcode} mass power spectrum models using Euclid-like survey configurations. Additionally, we introduce a new estimator to quantify scale leakage in both the BNT and noBNT approaches. Our findings show that BNT outperforms traditional methods, preserving cosmological constraints while significantly mitigating theoretical biases.
\end{abstract}

\maketitle

\section{Introduction}

A major achievement in cosmology over the past decades has been the development of the Standard Model of Cosmology (SMC) as a framework aimed at explaining the history of the universe and how the structure of the universe that we observe emerged. The SMC requires only seven parameters -- energy density for cold dark matter $\Omega_c$, baryon density $\Omega_b$, vacuum energy density $\Omega_\Lambda$, spectral index $n_s$, Hubble parameter $h$, $A_s$, and optical depth $\tau$  -- to fit a vast array of observational data, and these parameters are now known with remarkable precision \cite{2020A&A...641A...6P}. As data become increasingly precise, tensions have emerged in the measurement of some cosmological parameters across different probes \cite{2022JHEAp..34...49A}. As noted in \cite{2022JHEAp..34...49A,2022Univ....8..399D}, most of the anomalies and tensions involve data from the cosmic microwave background, suggesting that the late Universe ($z\sim 0-5$) appears slightly different from the early Universe ($z\sim 1100$). There is no consensus yet on whether these differences are due to residual systematics or new physics. This situation clearly calls for more data from the late Universe and a better understanding of the models used to interpret these data.

Weak Gravitational Lensing (WL) is the most reliable tool for probing mass distribution in the late Universe. By offering crucial insights into the expansion history and growth of instabilities of dark matter, it has the potential to address the nature of dark matter and dark energy and clarify the nature of these tensions \cite{2001PhR...340..291B}. 
The advent of Stage-IV lensing surveys, to be conducted by the European Space Agency's Euclid mission \cite{2011arXiv1110.3193L}, the Vera C. Rubin Observatory’s Legacy Survey of Space and Time \cite{2009arXiv0912.0201L}, and the Xuntian mission of the Chinese Space Station Telescope \cite{CSST} will greatly increase the statistical power of WL measurements while reducing residual systematics, offering a promising path to resolving current cosmological tensions.

It is important to remember that, unlike redshift surveys, WL is a two-dimensional probe—specifically, it measures the projected mass along lines of sight by correlating galaxy shapes as a function of their angular separation. The two-dimensional projection causes large-scale fluctuations from distant lenses to contribute to the signal at the same angular scale as smaller-scale fluctuations from nearby lenses. This is the case even if the redshifts of sources were known with precision, because all the lenses at lower redshift than the sources would still contribute to the WL signal. In Fourier space, this would lead to mode mixing: if we define $k$ as the modulus of the three-dimensional wavenumber of mass density fluctuations and $\ell$ as that of the two-dimensional wavenumber of the projected density fluctuations,  there is a wide range of $k$ modes at different redshifts $z$ which contribute to the same $\ell$ mode. This would not be an issue if the underlying model for the mass density fluctuations as a function of $k$ and $z$ was robust; however, this is not the case. The modeling of the mass distribution at small scale (high $k$) is particularly challenging. More specifically, the density fluctuations are determined by a combination of non-linear gravitational clustering and baryonic physics, both influenced by the complex processes of galaxy formation and evolution—none of which have a precise analytical description, especially for $k\gtrsim 0.1\;{\rm Mpc}^{-1}$. The modeling uncertainty can reach $10\%$ for $k\gtrsim 1\;{\rm Mpc}^{-1}$ and $30\%$ for $k\gtrsim 10\;{\rm Mpc}^{-1}$ \cite{2025PhRvL.134h0001C}. Additionally, the unknown nature of dark matter introduces further uncertainty regarding the statistical distribution of density fluctuations in large-scale structures\cite{2022arXiv220307354B}. 
The inability of the WL estimators to clearly separate contributions from lenses at different redshifts means that all these effects become mixed across all $\ell$'s.

This is a well-known limitation of WL, and currently, there are two ways to address this issue. One approach is to cross-correlate the WL signal with foreground lenses at known redshifts (e.g., using spectroscopic surveys). This technique, known as galaxy-galaxy lensing \cite{1993ApJ...404..441K} or cluster lensing \cite{1999ARA&A..37..127M}, has the drawback that the estimator is no longer an unbiased tracer of mass, and the interpretation of the signal depends on how the target lenses populate the underlying halo. The other approach is to exclude small angular scales (high $\ell$'s) from the cosmic shear analysis, since most issues are most pronounced at small physical scales (high $k$'s). However, the required scale cutoff is generally large (ten arcminutes or more), resulting in a significant loss of information, because high $\ell$ is not the same as high $k$. Adding tomography \cite{1999ApJ...522L..21H} -- using photometric redshift information as a proxy for distance -- does not fundamentally help, as lenses are still viewed in projection, even for a single source redshift slice. Scale cuts with tomography is the current approach for stage III surveys: Kilo Degree Survey \cite{2021A&A...645A.104A}, Dark Energy Survey \cite{2022PhRvD.105b3520A} and Hyper Suprime Cam Survey \cite{2023PhRvD.108l3521S,2023PhRvD.108l3519D}. This approach is not optimal because all lenses at redshift lower than the source redshift of a given tomographic bin still contribute to the WL signal of this tomographic bin. Consequently, the connection between physical scale and angular scale remains loose even with tomography. The question remains whether this approach can be improved for Stage IV surveys.

The BNT approach, \cite{2014MNRAS.445.1526B}, can provide an adequate framework to address this question, as already pointed in \cite{2018PhRvD..98h3514T,2021PhRvD.103d3531T}. The construction underlying the BNT transform aims at reorganizing the WL data vector in a way that separates the contribution of different physical scales. Its principle is to apply a linear transformation to the WL data such that the resulting lensing kernel is narrower than with the untransformed data. Consequently, the lenses of a BNT tomographic bin are distributed in a narrow redshift interval making the connection between physical scales and angular scales tighter. In harmonic space, the BNT transform reorganizes the elements of the WL data vector, the tomographic spectra, such that the new data vector is closer to three-dimensional $k$-space spectra. This approach offers two main advantages: (1) because the BNT transform is linear and invertible, it preserves all the information, and (2) with the BNT data arrangement, a cut in $\ell$-space is much closer to a cut in $k$-space. An $\ell$-space scale cut in the BNT rearranged data vector gives more control on what the WL estimator is sensitive to in $k$-space. For instance, as we will show in this paper, the low $\ell$ modes in the BNT data vector are much less sensitive to high $k$ modes than the same low $\ell$ modes taken from the noBNT data vector. Consequently, the BNT data vector offers a much cleaner probe of the fluctuation amplitude $S_8$. Alternatively, one might focus on the intermediate $\ell$ modes of the BNT data vector to investigate deviations in the mass power spectrum from predictions of Cold Dark Matter.

It is key to note that applying scale cuts to BNT transformed data is not equivalent to combined scale and photo-z cuts under a conventional tomographic approach. This will be the case for the scale cuts recipes advocated 
and used in this work.
The BNT transform is solely based on a property of the lensing kernel, recognizing that the WL signal from two different tomographic bins always share an identical subset of lenses, differing only by a purely geometric factor that is entirely independent of small-scale physics. This contribution from the subset can be reduced or even eliminated, for all tomographic bins. The BNT approach was initially introduced to extend the range of validity of standard Perturbation Theory results for cosmic shear observations. It considerably extends the reach of the nulling technique proposed in \cite{2005PhRvD..72d3002H,2008A&A...488..829J} introduced to mitigate the intrinsic alignment effect. In contrast, BNT reorganizes the data vector to suppress the WL signal from lenses located at lower redshifts than a given redshift bin. The number of tomographic bins becomes the limiting factor for BNT, giving Stage IV surveys a distinct advantage over Stage III surveys. In this paper, we demonstrate that BNT can effectively be used to make cosmic shear observations free of any assumptions on the behavior of the matter fluctuations at small scales.

The structure of the paper is as follows. In Section II, we compare the BNT and noBNT theories of WL and explain how the scale reorganization works. In Section III, we describe our methodology for the practical implementation of BNT. In Section IV, we present a forecast of parameter constraints using BNT and noBNT approaches, with the Euclid Stage IV survey as a benchmark, and demonstrate how various small-scale effects can be mitigated with BNT more efficiently than with conventional WL estimators. Finally, we summarize our findings and provide concluding remarks in the last section.

\section{Theory}

\subsection{Weak Lensing and BNT Transform}
\label{subsec:IIA}

The two-dimensional cosmic shear power spectrum, $C^{i,j}_\ell$, can be computed from the three-dimensional matter power spectrum $P(k;z)$ as follows \cite{1954ApJ...119..655L}:
\be
C^{i,j}(\ell) = \int\dfrac{d\chi}{\chi^2} W^i_\gamma(\chi)W^j_\gamma(\chi)P\left(k=\dfrac{\ell+1/2}{\chi};z(\chi)\right),
\label{Cell_original}
\ee
where $\chi$ is the radial comoving distance, and $W_\gamma^i(\chi)$ is the lensing kernel computed in the weak lensing regime of the $i^{th}$ tomographic bin of the given survey:
\be
W_\gamma^i(\chi) = \dfrac{\Omega_m^2H_0^4}{c^2}\int\dd \chi' \dfrac{n_i(\chi')}{a(\chi)}\dfrac{f_K(\chi'-\chi)f_K(\chi)}{f_K(\chi')}.
\ee

Here, $n_i(\chi)$ represents the redshift distribution of the source galaxies in the $i^{th}$ tomographic bin, $f_K(\chi)$ is the comoving angular diameter distance (which simplifies to $\chi$ in a flat universe), and $a(\chi)$ denotes the scale factor.

To construct the BNT transform matrix, we first define two normalisation numbers $n_i^{0}$ and $n_i^{1}$, for each tomographic bin $i$:
\bea
n_i^{0} &=& \int\dd\chi \; n_i(\chi) \\
n_i^{1} &=& \int\dd\chi \; \dfrac{n_i(\chi)}{\chi}
\eea

The BNT transform is aiming to reorganize the data vector $C^{i,j}(\ell)$ to a new one $\hat C^{a,b}(\ell)$, where $a$ and $b$ are new lensing kernels $\hat W_\gamma^a(\chi)$ and $\hat W_\gamma^b(\chi)$. Note that we will keep the `hat' notation $\hat{ }$ for BNT transformed quantities throughout the paper. We also reserve the use of $(i,j)$ for the noBNT tomographic bins and $(a,b)$ for the BNT tomographic bins. The new lensing kernels are linear combinations of the original ones:

\be
\hat W_\gamma^a(\chi) = \sum_{i=1}^{n_{\rm T}} p^a_i W_\gamma^i(\chi),
\label{BNTdef}
\ee
\begin{figure}[htbp]
\centering
\includegraphics[width=0.45\textwidth, trim = 0.1cm 4.35cm 0.1cm 0cm, clip]{BNT_matrix.png}
\caption{
The BNT transform matrix $p^a_i$ calculated with our fiducial cosmology and redshift distribution.
} \label{fig:BNT_pai}
\end{figure}
where $n_{\rm T}$ is the number of tomographic bins and $p^a_i$ are the BNT transform coefficients to be determined. We adopt the same approach as described in \cite{2014MNRAS.445.1526B}, where only three consecutive tomographic bins are used to define a new kernel $\hat W_\gamma^a(\chi)$ \footnote{This implies that implementing BNT requires a minimum of four tomographic bins.}. This means that the matrix $p^a_i$ is a $n_{\rm T}\times n_{\rm T}$ square matrix, with non-zero elements only in a diagonal band of width three (See Fig. \ref{fig:BNT_pai}). While it is theoretically possible to use more than three tomographic bins to construct $\hat W_\gamma^a(\chi)$, three bins is the solution which leads to the tightest connection between $k$ and $\ell$. 
Therefore, the $p^a_{i}$ coefficients are the solutions to the algebraic equations:
\bea
&\sum_{i=a-2}^a& p^a_{i} n_i^{0} = 0 \\
&\sum_{i=a-2}^a& p^a_{i} n_i^{1} = 0,
\eea
with $p^a_i = 0$ when $i \notin \{a-2, a-1, a\}$, $p_1^1 = p_2^2 = 1$, and $p_2^1 = -1$. The goal of the BNT transform is to localize each kernel $\hat W_\gamma^a(\chi)$ within its corresponding tomographic bin such that $\hat W_\gamma^a(\chi) \times \hat W_\gamma^b(\chi) = 0$ for $|a - b| \geq 2$. The other advantage of using three tomographic bins is that the solution $p^a_i$ can be derived analytically \cite{2014MNRAS.445.1526B}.
For a given 10-bin set of Euclid-like redshift distributions \cite{2020A&A...642A.191E}, such as the one shown in the upper panel of Figure \ref{fig:nz}, the corresponding weak lensing kernels $W_\gamma(z)$ and BNT-transformed lensing kernels $\hat W_\gamma(z)$ are presented in the middle panel of Figure \ref{fig:nz}. The BNT transform demonstrates a significant advantage by localizing the redshift dependence more effectively compared to the conventional lensing kernel. This advantage becomes even more pronounced when performing the 3D-to-2D projection, as illustrated in the lower panel of Figure \ref{fig:nz}.
It is important to note that the redshift distributions of the tomographic bins do not need to be non-overlapping; in fact, overlapping bins should be preserved to ensure an invertible transformation. Additionally, the first bin remains unchanged, while the second bin has its sensitivity to small-scale lenses reduced but not completely eliminated. This can be seen with the $p^a_i$ example calculated with our fiducial cosmology, shown on Figure \ref{fig:BNT_pai}. The BNT transformed angular power spectrum $\hat C^{a,b}(\ell)$ is given by

\bea
\hat C^{a,b}(\ell) &\equiv& \int \dfrac{\dd\ell}{\chi^2} \hat W_\gamma^a(\chi) \hat W_\gamma^b(\chi) P\left(\dfrac{\ell + 1/2}{\chi};z(\chi)\right) \nonumber\\
&=& p^a_ip^b_jC_\ell^{i,j},
\label{Cell_BNT}
\eea

The BNT transform fulfils two key principles \cite{2014MNRAS.445.1526B}:
\begin{itemize}
\item It is invertible, ensuring that no information is lost in the process.
\item It minimizes the overlap between the resulting kernels across different redshifts, enabling the isolation of information from distinct physical scales, based on a physical argument.
\end{itemize}

\begin{figure}[htbp]
\centering
\includegraphics[width=0.5\textwidth, trim = 1cm 1.5cm 2cm 2cm, clip]{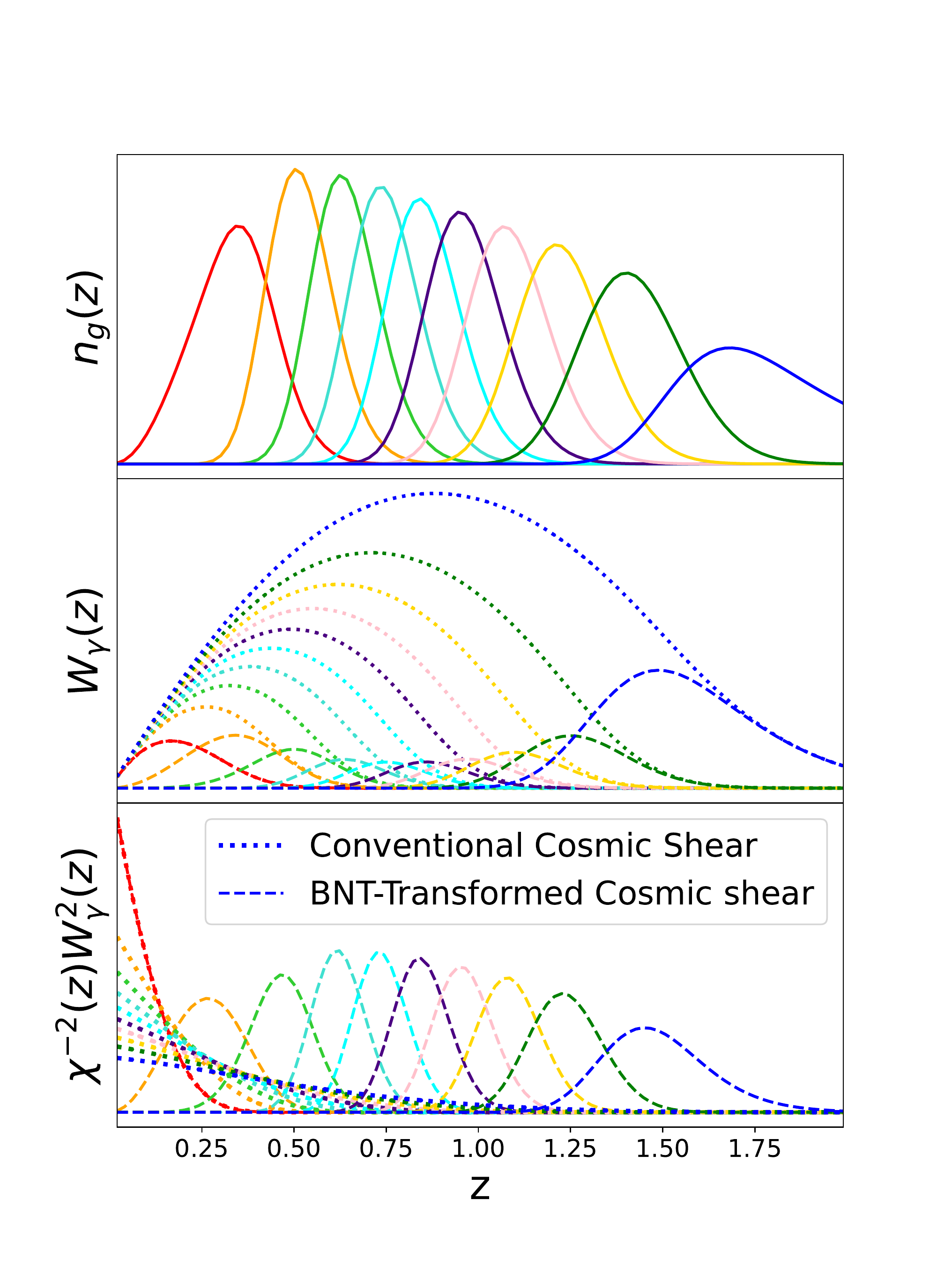}
\caption{Upper panel: Tomographic bins for the Euclid survey. Middle panel: conventional weak lensing kernels $W_\gamma^i(\chi)$ (dotted lines) and the BNT lensing kernel $\hat W_\gamma^a(\chi)$ (dashed lines). Lower panel: Lensing projection weights $W_\gamma^2(\chi)/\chi^2$ (dotted lines) and $\hat W_\gamma^2(\chi)/\chi^2$ (dashed ines) identifying the approximate location of the lenses for each kernel.
} \label{fig:nz}
\end{figure}

\subsection{Splitting in $k$-bins}
\label{subsec:IIB}

The BNT transform modifies the mapping between $k$-space and $\ell$-space. Therefore, it is crucial to understand how a $k$-bin, $[k_1, k_2]$, contributes to the two-dimensional power spectrum for both the BNT and noBNT cases. This is a key step to prepare the scale leakage discussion in Section \ref{subsec:IIIB}. The objective of this section is to introduce a notation for the data vector that allows us to identify the contribution of $k$-bins to the angular power spectrum.

We define $C^{i,j}_{[k_1,k_2]}(\ell)$ as the contribution to $C^{i,j}(\ell)$, for all values of $\ell$, from the $k$-bin $[k_1, k_2]$ of the three-dimensional mass power spectrum $P(k,z)$:

\bea \label{eqn:Cl_kcut}
C^{i,j}_{[k_1,k_2]}(\ell) &\equiv& \int_0^{\chi_\mathrm{H}} \frac{\dd \chi}{\chi^2} W_\gamma^i(\chi)W_\gamma^j(\chi) \mathbf{1}_{[k_1,k_2]}(k)  \cr
&\times & P\left(k,z(\chi)\right),
\label{eq:Celldef}
\eea
where $\mathbf{1}_{[k_1,k_2]}(k)$ is the top-hat function for the interval \([k_1,k_2]\) defined as:
\be
\mathbf{1}_{[k_1,k_2]}(k) =
\begin{cases}
1 & \text{if } k_1 < k < k_2, \\
0 & \text{otherwise}.
\end{cases} \label{fb:indicator}
\ee
with $k=\left(\frac{\ell+1/2}{\chi}\right)$. For the BNT case, the lensing kernel and the power spectrum are respectively replaced by $\hat{W}_\gamma^a(\chi)$ and $\hat{C}^{a,b}_{[k_1,k_2]}(\ell)$.
We define the data vector ${\cal C}_{[k_1,k_2]}(\Pi)$ as the collection of all tomographic bin combinations of $C^{i,j}_{[k_1,k_2]}(\ell;\Pi)$, where $\Pi$ is the set of cosmological parameters to constrain. For $n_{\rm T}=10$ tomographic bins, we have:
\be \label{eqn:tomo}
{\cal C}_{[k_1,k_2]}(\Pi) = \begin{pmatrix}
    \begin{array}{cc}
        \begin{bmatrix}
        C^{1,1}_{[k_1,k_2]}(\ell;\Pi)  \\
       \end{bmatrix} {\rm all}~\ell's \\
       \begin{array}{c}
       \end{array}\\
       \begin{bmatrix}
        C^{1,2}_{[k_1,k_2]}(\ell;\Pi) \\
       \end{bmatrix} {\rm all}~\ell's \\
       \begin{array}{c}
       \end{array}\\
       \begin{bmatrix}
        C^{1,3}_{[k_1,k_2]}(\ell;\Pi) \\
       \end{bmatrix} {\rm all}~\ell's \\
        ...& \\
       \begin{array}{c}
       \end{array}\\
       \begin{bmatrix}
        C^{9,10}_{[k_1,k_2]}(\ell;\Pi) \\
       \end{bmatrix} {\rm all}~\ell's \\
       \begin{array}{c}
       \end{array}\\
       \begin{bmatrix}
        C^{10,10}_{[k_1,k_2]}(\ell;\Pi) \\
       \end{bmatrix} {\rm all}~\ell's \\
    \end{array}
\end{pmatrix}.
\ee
The length of the ${\cal C}_{[k_1,k_2]}(\Pi)$ data vector is $\left[n_\ell \frac{n_{\rm T}(n_{\rm T}+1)}{2}\right]$, if there are $n_\ell$ is the number of $\ell$ bins. For the calculations, we restrict the $\ell$ space to $[\ell_{\rm min},\ell_{\rm max}]=[50,5000]$, where $50$ is the central value of the lowest bin.

In order to simplify the notation, we also define $\Delta {\cal C}_{[k_1,k_2]}(\Pi)$ as the difference between ${\cal C}_{[k_1,k_2]}(\Pi)$ and ${\cal C}_{[k_1,k_2]}(\Pi^0)$ where $\Pi^0$ is the fiducial set of cosmological parameters:

\be
\Delta {\cal C}_{[k_1,k_2]}(\Pi) \equiv  {\cal C}_{[k_1,k_2]}(\Pi)-{\cal C}_{[k_1,k_2]}(\Pi^0)
\label{Delta_Cdef}
\ee

The global data vector difference is defined as:

\begin{equation}
    \Delta {\cal D}(\Pi)=\sum_{[k_1,k_2]}\Delta {\cal C}_{[k_1,k_2]}(\Pi),
    \label{fb:deltaD}
\end{equation}
where the summation is performed over all $k$-bins, and by definition we have $\Delta {\cal D}(\Pi^0)=0$. The $\chi^2$ is defined as:

\begin{equation}
    \chi^2(\Pi)\equiv\Delta {\cal D}(\Pi)^T \cdot \mathbb{C}^{-1} \cdot \Delta {\cal D}(\Pi),\label{fb:chi2}
\end{equation}
where $\mathbb{C}$ is the $\left[n_\ell \frac{ n_{\rm T}(n_{\rm T}+1)}{2}\right] \times \left[n_\ell \frac{ n_{\rm T}(n_{\rm T}+1)}{2}\right]$ covariance matrix. For the BNT case, all quantities above are identified with a $\hat{}\;$ symbol, e.g. Eq. \ref{fb:chi2} becomes:

\begin{equation}
    \hat{\chi}^2(\Pi)\equiv\Delta {\cal \hat{D}}(\Pi)^T \cdot \mathbb{\hat{C}}^{-1} \cdot \Delta {\cal \hat{D}}(\Pi).
\end{equation}



\begin{figure}[htbp]
\centering
\includegraphics[width=0.45\textwidth, trim = 0.3cm 0.7cm 0.3cm 0.3cm, clip]{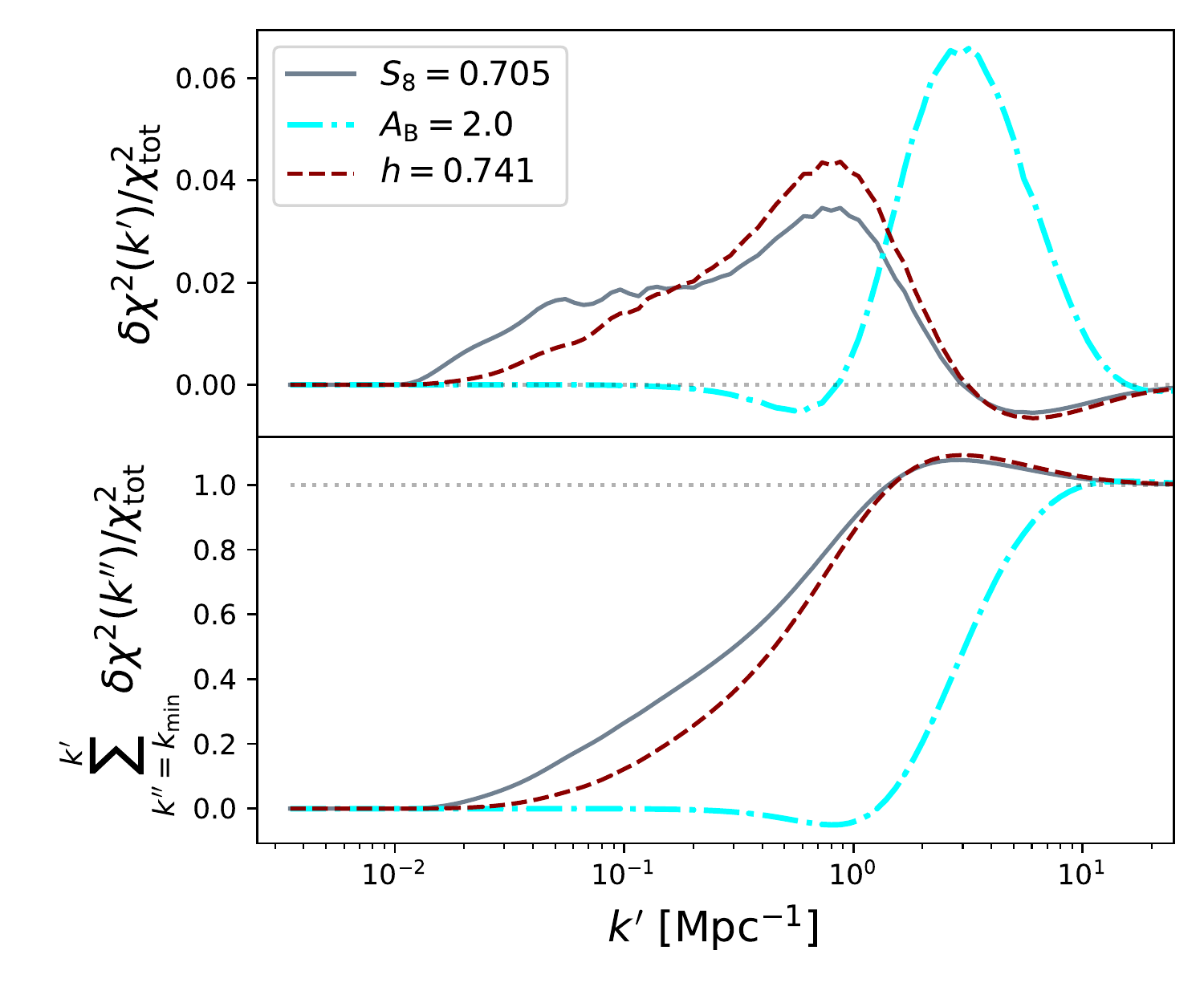}
\caption{$\chi^2_{\rm tot}$ is $\chi^2(\Pi^0)$ from Eq.\ref{chi_tot}, where the fiducial cosmology is given by $\Pi^0=(S_8,A_{\rm B},h)=(0.8124,3.13,0.690)$. Top panel: incremental contribution $\delta\chi^2_{[k_1,k_2]}(\Pi)$ from Eq.\ref{chi_incr}, where $k'$ is the center of $k$-bins $[k_1,k_2]$, for cosmology $\Pi=(0.705,3.13,0.690)$, solid line, $(0.8124,2.0,0.690)$, dot-dashed line and $(0.8124,3.13,0.741)$, dashed line. Bottom panel: cumulative contribution to $\chi^2_{\rm tot}$ from $k_{\rm min}= 0.0025\;\rm Mpc^{-1}$ to $k'$. The legend in the top panel specifies which of the three parameters has been modified relative to the fiducial values.}
\label{fig:dchi2}
\end{figure}

It is convenient to separate the contributions to Eq. (\ref{fb:chi2}) arising from the diagonal $k$-bins (autocorrelations) and the off-diagonal $k$-bins (cross-correlations). This can be accomplished by substituting Eq. (\ref{fb:deltaD}) into Eq. (\ref{fb:chi2}):

\begin{eqnarray}
        \chi^2(\Pi)&=&\left[\sum_{[k_1,k_2]}\Delta {\cal C}_{[k_1,k_2]}(\Pi)\right]^{^T} \cdot \mathbb{C}^{-1} \cdot \left[\sum_{[k'_1,k'_2]}\Delta {\cal C}_{[k'_1,k'_2]}(\Pi)\right]\nonumber\\
&=&\sum_{[k_1,k_2]}\Big[
\Delta {\cal C}^{^T}_{[k_1,k_2]}(\Pi) \cdot \mathbb{C}^{-1}\cdot \Delta {\cal C}_{[k_1,k_2]}(\Pi)\Big] \nonumber \\
&+&\sum_{[k_1,k_2]}\left[
\Delta {\cal C}^{^T}_{[k_1,k_2]}(\Pi) \cdot \mathbb{C}^{-1} \cdot \sum_{\substack{[k'_1,k'_2]\ne \\  [k_1,k_2]}}\Delta {\cal C}_{[k'_1,k'_2]}(\Pi)
\right] \nonumber\\
&\equiv&\sum_{[k_1,k_2]}\left[\delta\chi^2_{[k_1,k_2],\mathrm{auto}}(\Pi)+\delta\chi^2_{[k_1,k_2],\mathrm{cross}}(\Pi)\right].
\label{chi2_sum}
\end{eqnarray}

which can be rewritten as:
\be
\chi^2(\Pi) = \sum_{[k_1,k_2]}\delta\chi^2_{[k_1,k_2]}(\Pi),
\label{chi_tot}
\ee

where $\delta\chi^2_{[k_1,k_2]} $ is defined as the incremental contribution to $\chi^2(\Pi)$ originating from the scale bin $[k_1,k_2]$:
\be
\delta\chi^2_{[k_1,k_2]}(\Pi) \equiv \delta\chi^2_{[k_1,k_2],\mathrm{auto}}(\Pi) + \delta\chi^2_{[k_1,k_2],\mathrm{cross}}(\Pi).
\label{chi_incr}
\ee

$\delta\chi^2_{[k_1,k_2]}(\Pi)$ is a useful metric for estimating the contribution of each $k$-bin to $\chi^2(\Pi)$. Note that the cross term $\delta\chi^2_{[k_1,k_2],\mathrm{cross}}$ incorporates information from outside the $[k_1,k_2]$ bin through its cross-correlation with modes within $[k_1,k_2]$.
Figure~\ref{fig:dchi2} illustrates this concept with examples involving the lensing degeneracy parameter $S_8$, the Hubble constant $h$, and the baryon feedback parameter $A_\mathrm{B}$ \cite{2001MNRAS.321..559B} from \textsc{HMcode} \cite{HMcode16}, which describes the halo mass concentration. The plot reveals different sensitivity to the $k$ bin range for different parameters: both the Hubble constant and $S_8$ exhibit peak sensitivity at $k\sim0.9$ Mpc$^{-1}$, while $A_\mathrm{B}$ peaks at $k\sim3$ Mpc$^{-1}$.
Interestingly, the examples show a negative contribution to $\delta\chi^2$ at certain scales for all three parameters. This behaviour occurs only when nonlinear terms are included in the covariance matrix calculation, indicating the presence of negative correlations in the cross-band correlations $\delta\chi^2_{\rm cross}$ between different $k$-modes. These negative contributions are associated with the off-diagonal terms of the covariance matrix. We also found that the $k$ bin sensitivity does not significantly differ between the BNT and non BNT cases for the original binning setup in $\ell$-space.

Following the notation introduced in Section \ref{subsec:IIA}, the BNT version of all the expressions of this section are obtained with the $\hat{}\;$ symbol of each quantity. For instance, ${\cal C}_{[k_1,k_2]}(\Pi)$ is the noBNT data vector (Eq.\ref{eqn:tomo}), while the BNT version is $\hat{{\cal C}}_{[k_1,k_2]}(\Pi)$ is similarly described by Eq.\ref{eqn:tomo}, with the $C_{[k_1,k_2]}^{i,j}(\ell;\Pi)$ (Eq.\ref{Cell_original}) replaced by $\hat{C}_{[k_1,k_2]}^{a,b}(\ell;\Pi)$ (Eq.\ref{Cell_BNT}). The practical calculation of Eqs. \ref{chi_tot} and \ref{chi_incr} are given in Appendix \ref{appendix:deltachi2}.

\section{Optimizing scale cuts}

\subsection{Methodology}
\label{subsec:IIIA}

\begin{figure*}[htbp]
\centering
\includegraphics[width=0.95\textwidth, trim = 6cm 6cm 6cm 6cm, clip]{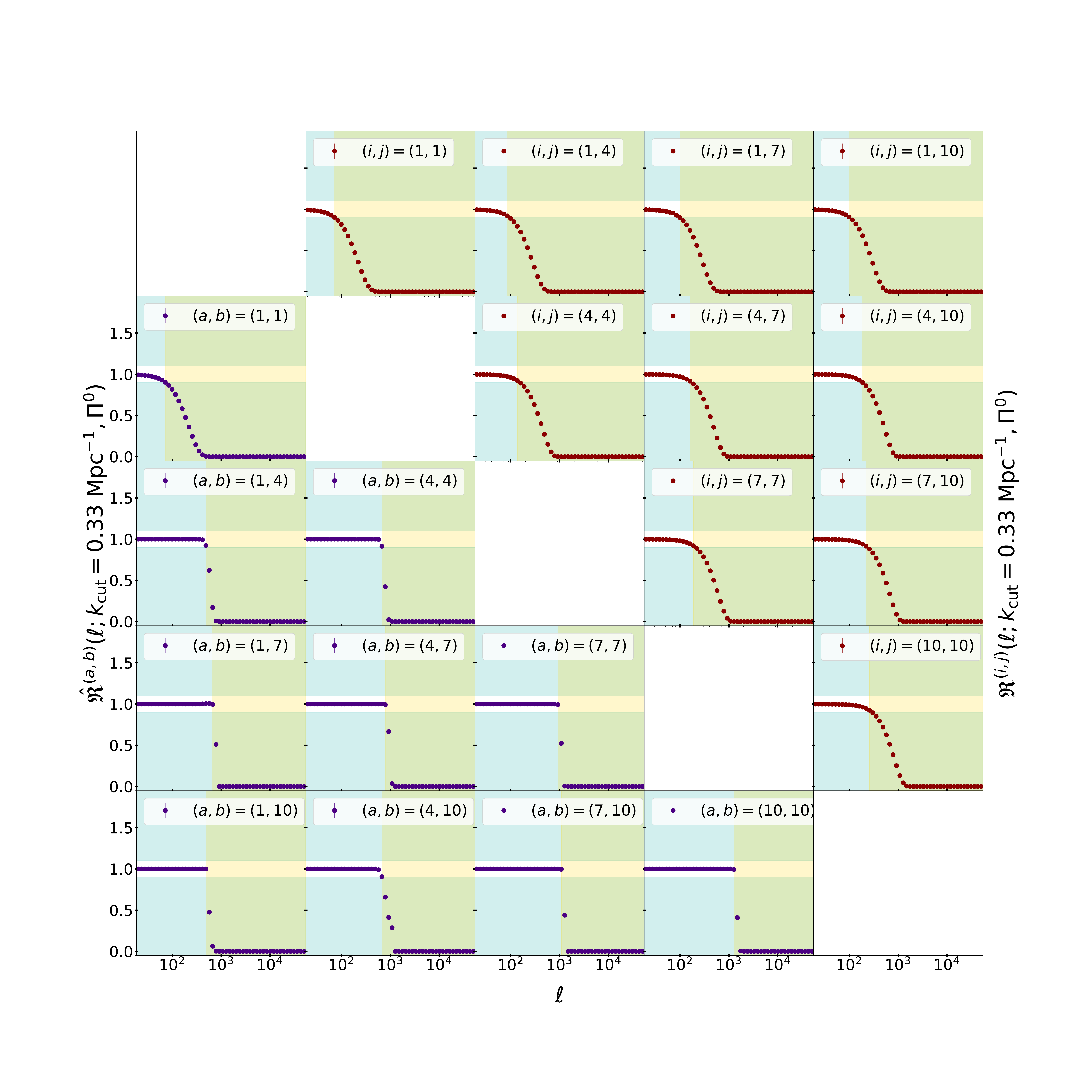}
\caption{Ratio of the data vector with $k_{\rm cut}$ to the full data vector (defined in Eq. \ref{Rfrak_def}) for the noBNT case $\fR^{(i,j)}(\ell;k_{\rm cut},\Pi^0)$ (upper-right triangle) and the BNT case $\hfR^{(i,j)}(\ell;k_{\rm cut},\Pi^0)$ (lower-left triangle). Each panel shows the ratio for the combination of tomographic bins indicated by $(i,j)$ (for the noBNT case) and $(a,b)$ (for the BNT case). For each tomographic bin, the curves are used to define which cut in $\ell$ corresponds to the constraint that modes $k>k_{\rm cut}$ should not bias the data vector at a level exceeding $\mathcal{T}_\mathrm{FD}$ for $\ell<\ell_{\rm cut}^{(i,j)}$ for noBNT (resp. $\ell<\hat{\ell}_{\rm cut}^{(a,b)}$ for BNT). All plots are given for $k_\mathrm{cut} = 0.33\;{\rm Mpc^{-1}}$, and the horizontal bright band indicates the fractional threshold $\mathcal{T}_\mathrm{FD} = \pm 0.1$. For each tomographic bin combination, the cut in $\ell$ is shown as the boundary of the cyan and green regions.
}
\label{fig:Cl}
\end{figure*}

In this section, we will demonstrate why a cut in $\ell$-space for the BNT transform corresponds more closely to a cut in $k$-space compared to the noBNT case. To quantify the relationship between $k$ and $\ell$ cuts for the BNT and noBNT cases, the following dimensionless ratios are introduced:

\bea
\hfR^{(a,b)} (\ell;k_{\rm cut},\Pi^0) &\equiv& \frac{\hmC^{a,b}_{[k_\mathrm{min},k_\mathrm{cut}]}(\ell;\Pi^0)}{\hmC^{a,b}_{[k_\mathrm{min},k_\mathrm{max}]}(\ell;\Pi^0)} \nonumber\\
{\fR}^{(i,j)}(\ell;k_{\rm cut},\Pi^0)&\equiv&\frac{{\cal C}^{i,j}_{[k_\mathrm{min},k_\mathrm{cut}]}(\ell;\Pi^0)}{{\cal C}^{i,j}_{[k_\mathrm{min},k_\mathrm{max}]}(\ell;\Pi^0)}
\label{Rfrak_def}
\eea

We choose the fiducial cosmology $\Pi^0$ to perform the BNT transform; however, it is important to emphasize that this choice does not affect the results. The BNT transform is not unique—it can be any linear transformation of the original data vector entries. Consequently, there is no `exact' BNT transform, and the selection of a specific cosmology to calculate the $p_i^a$ is inconsequential, it does not have to be the {\it right} one. The only consequence of not using the "right" cosmology is that non overlapping BNT redshift kernels might slightly overlap, resulting in an imperfect nulling (which can be exploited on its own as stressed in \cite{2025arXiv250202243T,2025arXiv250202246T}), however, our approach remains unbiased even with imperfect nulling. The cosmological parameters are now known with such high precision that the choice of a particular cosmology within the current uncertainties has a negligible impact. 
We simply choose a cosmology that we believe is reasonably close to the actual cosmological parameters of the universe. 

Figure \ref{fig:Cl} shows $\hfR^{(a,b)} (\ell;k_{\rm cut},\Pi^0)$ and ${\fR}^{(i,j)}(\ell;k_{\rm cut},\Pi^0)$ for $k_{\rm cut}=0.3\;{\rm Mpc^{-1}}$. Plotting these ratios as a function of $\ell$ for a specific $k_{\rm cut}$ illustrates the impact of a cut in $k$-space on the angular power spectrum. This analysis enables the definition of an $\ell$-space cut that closely approximates the effect of a cut in $k$-space. For small enough $\ell$, all ratios converge to unity, indicating that almost all the signal is coming from $k\in [k_\mathrm{min},k_\mathrm{cut}]$. For high enough $\ell$, all ratios converge to zero, indicating that almost no signal is coming from $k\in [k_\mathrm{min},k_\mathrm{cut}]$. The BNT ratios $\hfR^{(a,b)} (\ell;k_{\rm cut},\Pi^0)$ show that the transition between low and high $\ell$ is much sharper than for the noBNT ratios ${\fR}^{(i,j)}(\ell;k_{\rm cut},\Pi^0)$. Figure \ref{fig:Cl} shows that a $k$-space cut impacts many more $\ell$ modes in the noBNT data vector than the BNT one. This illustrates why a cut in $\ell$-space is much closer to a cut in $k$-space with BNT. It is particularly striking that for noBNT, even very low $\ell$ are affected by a cut at $k_{\rm cut}=0.3\;{\rm Mpc^{-1}}$. 

Figure \ref{fig:Cl} shows why the $\ell$ cut value must depend on the tomographic combination. We define  $\hat\ell^{(a,b)}_{\rm cut}$ as the $\ell$ cuts for BNT and $\ell^{(i,j)}_{\rm cut}$ for noBNT. We also define a $\tFD$ the threshold as a tolerance region with $\hfR^{(a,b)} (\ell=\hat\ell^{a,b}_{\rm cut};k_{\rm cut},\Pi^0)=1 \pm \tFD$.
Figure \ref{fig:Cl} shows how $\hat\ell^{(a,b)}_{\rm cut}$ (resp. $\ell^{(i,j)}_{\rm cut}$) are chosen for $k_{\rm cut}=0.3\;{\rm Mpc^{-1}}$ and $\mathcal{T}_\mathrm{FD} = 0.1$, for the BNT (resp. noBNT) cases. The yellow area shows the $\pm \mathcal{T}_\mathrm{FD}$ region for $\hfR^{(a,b)}$ (resp. $\fR^{(i,j)}$). The blue area shows the range $\ell < \hat\ell^{(a,b)}_{\rm cut}$ (resp. $\ell < \ell^{(i,j)}_{\rm cut}$) meets the $\mathcal{T}_\mathrm{FD}$ criteria and the green area shows the range of $\ell$ that must be removed in order to meet the criteria.
It is worth noting that this approach could, in principle, be extended to the shear-shear correlation functions $\xi_\pm$ and the COSEBIs $W_n(\ell)$. However, these summary statistics introduce additional scale mixing, beyond the simple two-dimensional projection, making the BNT approach less effective for these cases.

The selection of specific values for $k_{\rm cut}$ and $\mathcal{T}_\mathrm{FD}$ is somewhat arbitrary. However, it is evident that for a given $k_{\rm cut}$, increasing $\mathcal{T}_\mathrm{FD}$ incorporates more information from higher $k$-modes, which is contrary to the purpose of the BNT method. The ideal scenario would be to make $\mathcal{T}_\mathrm{FD}$ sufficiently small to ensure that cosmological parameter measurements are insensitive to scales $k > k_{\rm cut}$. In the following, we use this formalism to evaluate the performance of a particular $(k_{\rm cut}, \mathcal{T}_\mathrm{FD})$ configuration in terms of $k$-space leakage (i.e., how much scales $k > k_\mathrm{cut}$ contribute). We also show that, compared to the noBNT approach, significant information can still be retained with BNT, even for very small $\mathcal{T}_\mathrm{FD}$ values.

In order to incorporate the effect of $(k_\mathrm{cut},\mathcal{T}_\mathrm{FD})$ in the data vector, it is convenient to define the continuous Boolean weight function ${\mathcal{V}}(k_\mathrm{cut},\mathcal{T}_\mathrm{FD})$:

\be \label{eqn:ellcut}
{\mathcal{V}}(k_\mathrm{cut},\mathcal{T}_\mathrm{FD}) = \begin{cases}
1, |\fR (k_{\rm cut},\Pi^0)-1| \leq \mathcal{T}_\mathrm{FD} \\
0, |\fR (k_{\rm cut},\Pi^0)-1| > \mathcal{T}_\mathrm{FD},
\end{cases}
\ee
where $\fR (k_{\rm cut},\Pi^0)$ is the power spectra ratio for the entire data vector Eq.\ref{eqn:tomo} in the noBNT case \footnote{For the BNT case, all quantities are identified with a hat, e.g. $\hat{\mathcal{V}}(k_\mathrm{cut},\mathcal{T}_\mathrm{FD})$, $\hfR (k_{\rm cut},\Pi^0)$}:

\be
\fR (k_{\rm cut},\Pi^0) \equiv \frac{{\cal C}_{[k_\mathrm{min},k_\mathrm{cut}]}(\Pi^0)}{{\cal C}_{[k_\mathrm{min},k_\mathrm{max}]}(\Pi^0)}
\ee

We can rewrite Eq.\ref{fb:chi2} as

\be
\chi^2(\Pi;k_\mathrm{cut},\mathcal{T}_\mathrm{FD})=\Delta {{\cal D}^*}^T \cdot (\mathbb{C^*})^{-1} \cdot \Delta {\cal D}^*,\label{fb:chi2kFD}
\ee
where the $^*$ quantities refer to the global data vector (Eq.\ref{fb:deltaD}) and covariance matrix truncated with the boolean filter Eq. \ref{eqn:ellcut}:

\bea
\Delta {\cal D}^* &\equiv& \mathcal{V}(k_\mathrm{cut},\mathcal{T}_\mathrm{FD})\Delta {\cal D}(\Pi) \nonumber\\
\mathbb{C^*} &\equiv& \mathcal{V}(k_\mathrm{cut},\mathcal{T}_\mathrm{FD}) \cdot \mathbb{C} \cdot \mathcal{V}(k_\mathrm{cut},\mathcal{T}_\mathrm{FD})^{^T}
\label{star_eqn}
\eea

\begin{figure*}[htbp]
\centering
\includegraphics[width=0.98\textwidth, trim = 1.2cm 1.7cm 2.8cm 2.2cm, clip]{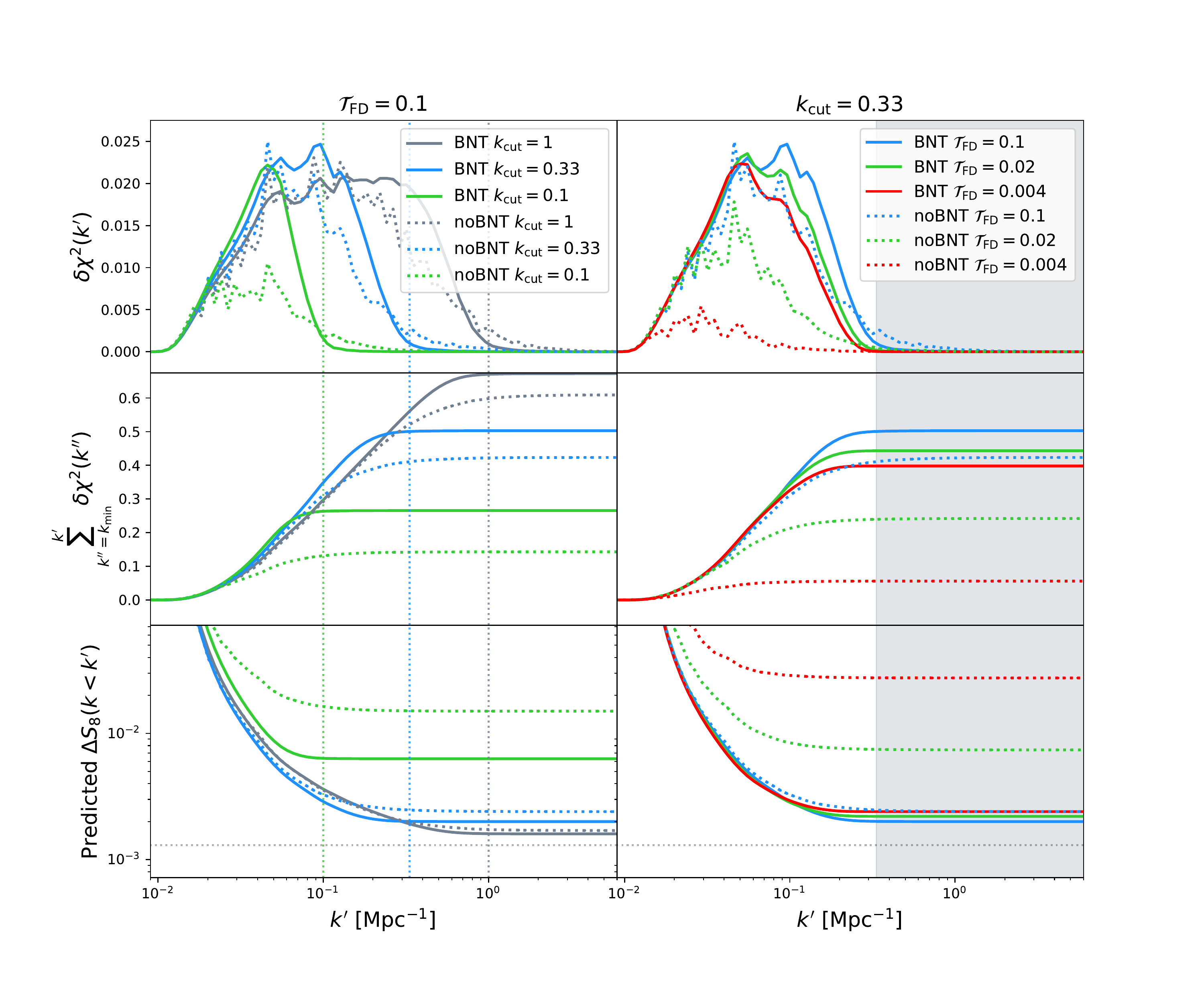}
\caption{All curves use $S_8=0.705$, while the fiducial cosmology $\Pi^0$ is the same as in Figure \ref{fig:dchi2} with $S_8=0.8124$. Dotted curves are for noBNT and solid curves are for BNT. The meaning of the different line colors is indicated in the legend box in the top two panels. The top two panels show the incremental $\delta\chi^2(k')$ (Eq.\ref{chi_incr}) as a function of $k'$, the midpoint of a $[k_1,k_2]$ interval. The middle panels show the cumulative $\delta\chi^2(k'')$, where $k''$ is summed from $k_{\rm min}$ to $k'$. The bottom panels show the predicted errorbar for $S_8$. All panels on the left are for $\mathcal{T}_\mathrm{FD} = 0.1$ and $k_{\rm cut}=[0.1,0.33,1]\;\rm Mpc^{-1}$. The vertical grey dotted lines mark the three $k_{\rm cut}$ values. All panels on the right are for $k_{\rm cut}=0.33\;{\rm Mpc^{-1}}$ and $\mathcal{T}_\mathrm{FD}=[0.004,0.02,0.1]$. The grey region in the right panels indicates modes $k'>k_{\rm cut}=0.33\;{\rm Mpc^{-1}}$.
} \label{fig:deltachi2}
\end{figure*}

We can now rewrite all equations in Sections \ref{subsec:IIB} with the filtered global data vector $\Delta {\cal D}^*$ and covariance matrix $\mathbb{C^*}$. In particular, there is a filtered version of Eq.\ref{DeltaC_approx}:

\be
\Delta{\cal C}^{0^*}_{[k_1,k_2]}(\Pi) \equiv {\cal W}_{[k_1,k_2]}(\Pi^0)\Delta{\cal D}^*(\Pi)
\ee

The formalism for comparing the performance of BNT and noBNT analyses is now fully established. The procedure, being the same for noBNT and BNT, is as follows: for a given set of parameters $(k_{\rm cut}, \mathcal{T}_\mathrm{FD})$, the angular scale cuts $\ell_{\rm cut}^{(i,j)}$ (resp. $\hat{\ell}_{\rm cut}^{(a,b)}$) using the criteria defined by Eq. \ref{eqn:ellcut}. The $\chi^2$ (Eq.\ref{chi2_sum}) and the incremental $\delta\chi^2$ (Eq.\ref{chi_incr}) are calculated using the global data vector and covariance matrix (Eq. \ref{star_eqn}) truncated by the boolean filter $\mathcal{V}(k_\mathrm{cut}, \mathcal{T}_\mathrm{FD})$ (resp. $\hat{\mathcal{V}}(k_\mathrm{cut}, \mathcal{T}_\mathrm{FD})$). For practical implementation details, we direct the reader to Appendix \ref{appendix:deltachi2}, specifically Eqs. \ref{fb:dchi2auto}, \ref{fb:dchi2res}, and \ref{eqn:dchi2}.

It is worth revisiting the constraining power of various $k$-bins when using the filtered data vector, previously illustrated in Figure \ref{fig:dchi2} with the original data vector. Figure \ref{fig:deltachi2} presents calculations of $\delta\chi^2$ (Eq.\ref{chi_incr}). The left column corresponds to $\mathcal{T}_\mathrm{FD}=0.1$ and $k_{\rm cut}=[0.1,0.33,1]\;\rm Mpc^{-1}$, while the right column corresponds to $k_{\rm cut}=0.33\;{\rm Mpc^{-1}}$ and $\mathcal{T}_\mathrm{FD}=[0.004,0.02,0.1]$. The fiducial cosmology $\Pi^0$ remains the same as in Figure \ref{fig:dchi2} (i.e., $S_8=0.8124$), while in Figure \ref{fig:deltachi2}, only $S_8$ is modified to $0.705$.

The top two panels of Figure \ref{fig:deltachi2} display the incremental $\delta\chi^2(k')$ as a function of $k'$, the midpoint of a $[k_1,k_2]$ bin. The larger the area under the $\delta\chi^2(k')$ curves, the greater the constraining power. It is evident that the BNT method systematically provides more constraining power than the noBNT approach. Additionally, the maximum $\delta\chi^2$ remains significant even for $\mathcal{T}_\mathrm{FD}=0.004$ or $k{\rm cut}=0.1\;{\rm Mpc^{-1}}$. The key takeaway from the top two panels is that although the constraining power drops to zero at $k'\gtrsim k_{\rm cut}$ rather than precisely at $k_{\rm cut}$, this decline is much steeper and less dependent on $k'$ for BNT than for noBNT. The top-right panel illustrates scale leakage: the extent to which the curve extends into the grey region (starting at $k_{\rm cut}=0.33\;{\rm Mpc^{-1}}$ indicates the contribution of modes with $k > k_{\rm cut}$ to a given $\ell_{\rm cut}$. This relationship is determined by the choice of $(\mathcal{T}_\mathrm{FD}, k{\rm cut})$. A more detailed discussion of scale leakage is provided in the next subsection.

The middle panels of Figure \ref{fig:deltachi2} display the cumulative $\delta\chi^2(k')$, illustrating how $\chi^2$ asymptotically approaches its final value as $k' \rightarrow k_\mathrm{cut}$. 
The derivative of the cumulative $\delta\chi^2(k')$ is steeper for BNT compared to noBNT, further highlighting the issue of scale leakage.
Notably, for a fixed $k_{\rm cut}$, the asymptotic values of $\chi^2$ remain relatively consistent across different $\mathcal{T}_\mathrm{FD}$ values for BNT, while they vary significantly for noBNT. This behaviour demonstrates that noBNT quickly loses its constraining power when $\mathcal{T}_\mathrm{FD}$ is significantly reduced. This occurs because the constraining power of noBNT is dominated by mixed scales, whereas for BNT, the constraining power is better optimized for the chosen $k_{\rm cut}$.

The bottom panels of Figure \ref{fig:deltachi2} illustrate how the error on $S_8$ scales with $k'$ as more $k$-bins are included in the $\chi^2$, based on the approximation that the error bar on $S_8$ is inversely proportional to $\chi^2$. A comparison between the solid and dotted curves demonstrates that for $\mathcal{T}_\mathrm{FD}=0.004$, the BNT approach significantly outperforms the noBNT approach by a factor of approximately 10. Additionally, the bottom-left panel shows that reducing $k_{\rm cut}$ has a similar effect to decreasing $\mathcal{T}_\mathrm{FD}$.

\subsection{Scale Leakage}
\label{subsec:IIIB}

\begin{figure*}[htbp]
\centering
\includegraphics[width=1.0\textwidth]{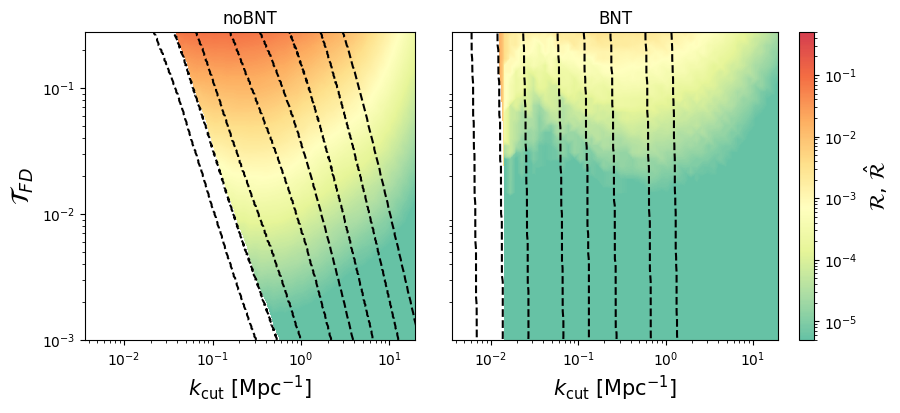}
\caption{Colourmap of the ratio $\mathcal{R}$ (Eq. \ref{ratioR2}) for the noBNT (left panel) and $\mathcal{\hat{R}}$ for BNT (right panel) as a function of $k_{\rm cut}$ and $\mathcal{T}_\mathrm{FD}$, characterizing the leakage. The white region on the left side of each panel corresponds to the limit $\ell^{10,10}_{\rm cut}\simeq 50$ ($\theta \simeq 220'$) for the most distant tomographic bin combination, when all $\ell$ bins drop below our lowest bin $\ell_{\rm min}=50$. The black dashed lines on each panel indicate (from left to right) the boundary of the white region if our lowest bin was $\ell_{\rm min} = 25, 50, 100, 250, 500, 1000, 2500$ and $5000$. The dashed lines can be used to identify which maximum $\ell$ is accessible from observations, for a given $k_{\rm cut}$ and $\mathcal{T}_\mathrm{FD}$.
}
\label{fig:color_BNT}
\end{figure*}

Scale leakage refers to the contribution of the $k > k_{\rm cut}$ modes to a data vector restricted to $\ell < \ell_{\rm cut}$. Quantifying scale leakage is essential as it provides insights into which regions of the mass power spectrum can be trusted, the level of precision achievable, and the impact on cosmological parameter measurements. For instance, \cite{2021ApJ...908...13Y} and \cite{2023MNRAS.525.4871G} demonstrated that a biased model for the nonlinear power spectrum, such as one influenced by baryonic physics, could lead to a tension in $S_8$.
Figure \ref{fig:dchi2} shows that the parameter $A_\mathrm{B}$ \cite{2001MNRAS.321..559B} can specifically bias the power spectrum for $k > 1\;$Mpc$^{-1}$, potentially resulting in a biased $S_8$ if left uncorrected. Marginalizing over $A_\mathrm{B}$ could mitigate this issue; however, the underlying mass power spectrum model could still be biased, and the associated loss of information could be significant. When using a BNT-reordered data vector, understanding scale leakage becomes even more critical to properly select the cuts $\hat{\ell}^{(a,b)}_{\rm cut}$ in accordance with the range of $k$ modes where the 3-dimensional power spectrum can be trusted, i.e. with a bias below a certain level. 

Scale leakage was negligible for Stage I and II surveys, where small scale physics was largely dominated by the total error budget due to the limited survey areas and residual systematics. While scale leakage may contribute to Stage III surveys, for instance it could partially explain the $S_8$ tension with the Cosmic Microwave Background, \cite{2023MNRAS.525.4871G,2025arXiv250217339P}, implementing BNT for Stage III remains challenging due to the limited statistical power and photometric redshift precision. In contrast, as we will demonstrate in the remainder of this work, scale leakage plays a crucial role in Stage IV surveys, where Stage IV observing conditions are considered.

In the following, we will use $\ell_{\rm cut}$ to denote any tomographic dependent cut in $\ell$, either $\hat{\ell}^{(a,b)}_{\rm cut}$ for BNT or $\ell^{(i,j)}_{\rm cut}$ for noBNT, in order to simplify the notation. With this notation, $\ell<\ell_{\rm cut}$ means that the cuts in $\ell$ are different for each tomographic bin $\ell<\ell^{(i,j)}_{\rm cut}$ (resp. $\ell<\hat{\ell}^{(a,b)}_{\rm cut}$), as explained in section \ref{subsec:IIIA}.
The scale leakage is defined as the ratio $\cal R$:

\be
\mathcal{R} \equiv \frac{\chi^2[k > k_\mathrm{cut}, \ell < \ell_\mathrm{cut}]}{\chi^2[\ell < \ell_\mathrm{cut}]},
\label{ratioR}
\ee
where the numerator is given by:

\be
\chi^2[k > k_\mathrm{cut}, \ell < \ell_\mathrm{cut}]=\Delta {{\cal D}^\dagger}^{^T} \cdot (\mathbb{C^*})^{-1} \cdot \Delta {\cal D}^\dagger,\label{eqn:chi2_conflicted}
\ee

with $\Delta {\cal D}^\dagger(\Pi)$ defined as the part of the global data vector satisfying both conditions $k > k_\mathrm{cut}$ and $\ell < \ell_\mathrm{cut}$:

\be
\Delta {\cal D}^\dagger(\Pi) \equiv \mathcal{V}(k_\mathrm{cut},\mathcal{T}_\mathrm{FD})\left[\sum_{[k_1,k_2]>k_{\rm cut}}  \Delta {\cal C}_{[k_1,k_2]}(\Pi)\right].
\ee

The denominator of Eq. \ref{ratioR} is given by Eq. \ref{eqn:ellcut} and Eq. \ref{fb:chi2kFD}, referred to as $\chi^2[\ell<\ell_{\rm cut}]$. Eq. \ref{ratioR} is the fraction of $k > k_\mathrm{cut}$ modes contributing to the total $\chi^2$ for a given $\ell_{\rm cut}$.

In order to illustrate the impact of scale leakage on $S_8$, with and without BNT, we adopt the same approach as \cite{2022MNRAS.516.5355A}, where the deviation of the fiducial cosmological model in the non-linear scales was described by a new parameter $A_{\rm mod}$:

\be
P(k,z)\equiv P^{\rm L}(k,z)+A_{\rm mod}(P^{\rm NL}(k,z)-P^{\rm L}(k,z)),
\ee
where $P^{\rm L}$ and $P^{\rm NL}$ represent the linear and nonlinear power spectra, respectively. To compare the scale leakage in the noBNT and BNT cases, Eq. \ref{ratioR} is calculated using $(A_{\rm mod}, S_8) = (0.827, 0.767)$ with $(A_{\rm mod}, S_8)_{\rm fid} = (1.0, 0.8124)$ for the fiducial model \footnote{The particular choice of $(A_{\rm mod}, S_8) = (0.827, 0.767)$ is motivated by the measurement $S_8=0.766^{+0.020}_{-0.014}$ \cite{2021A&A...646A.140H}, and a one-parameter minimization process of $A_{\rm mod}$ leading to $A_{\rm mod}=0.827$. This value is obtained by minimizing $\chi^2$ with $S_8 = 0.767$ and keeping other parameters unchanged.}.

To quantify the scale leakage from the nonlinear regime in relation to the $S_8$ tension, we adopt the following strategy: the denominator of Eq. \ref{ratioR} is calculated using $(A_{\rm mod}, S_8) = (0.827, 0.767)$, while the numerator accounts for the contributions of $S_8$ and $A_{\rm mod}$ separately:

\be
\mathcal{R} = \frac{\chi^2_{S_8}[k > k_\mathrm{cut}, \ell < \ell_\mathrm{cut}]+\chi^2_{A_\mathrm{mod}}[k > k_\mathrm{cut}, \ell < \ell_\mathrm{cut}]}{\chi^2_{S_8,A_\mathrm{mod}}[\ell < \ell_\mathrm{cut}]},
\label{ratioR2}
\ee
where the subscript $S_8$ or $A_{\rm mod}$ indicates which parameter deviates from its fiducial value.

The ratio $\cal R$ is shown in Figure \ref{fig:color_BNT} across a wide range of $k_{\rm cut}$ and $\mathcal{T}_\mathrm{FD}$. A low $\cal R$ indicates minimal scale leakage.
Figure \ref{fig:color_BNT} demonstrates that, for a given $k_{\rm cut}$, BNT exhibits significantly less scale leakage (i.e. $\mathcal{\hat{R}}\ll 1$) than noBNT. Similarly, for a fixed $\mathcal{T}_\mathrm{FD}$, BNT systematically shows less leakage than noBNT, and the minimum $k_{\rm cut}$ attainable by BNT is significantly smaller than for noBNT. Overall, $k$-mode mixing is far more pronounced for noBNT than for BNT. Consequently, for noBNT, using a low $\mathcal{T}_\mathrm{FD}$ necessitates a high $k_{\rm cut}$ in order to minimize the scale leakage, resulting in substantial information loss compared to BNT. The right panel of Figure \ref{fig:color_BNT} illustrates why an $\ell_{\rm cut}$ in the BNT case closely aligns with a $k_{\rm cut}$, because the iso-$\ell$ lines are nearly aligned with constant $k_{\rm cut}$.
The white region on the left side of each panel indicates the limit where the data vector is empty, as a consequence that $\ell$ below our lowest bin $\ell_{\rm min}=50$ are not included. This occurs when $\ell_{\rm cut}^{(10,10)} < \ell_{\rm min}$, where $\ell_{\rm cut}^{(10,10)}$ is the highest redshift bin combination for $n_T=10$. For noBNT, $\ell_{\rm cut}^{(10,10)}$ is a strong function of $\mathcal{T}_\mathrm{FD}$ and $k_{\rm cut}$, for BNT it is nearly independent of $\mathcal{T}_\mathrm{FD}$, which is another illustration that, for BNT, an $\ell$ and $k$ cuts have a similar impact on the data vector. Dashed lines in Figure \ref{fig:color_BNT} shows where the white region would start for a different $\ell_{\rm cut}^{(10,10)}$ value. One can also understand these lines as showing the largest possible $\ell$ accessible from observations, for a given $k_{\rm cut}$ and $\mathcal{T}_\mathrm{FD}$. With BNT, we can still take advantage of $\ell \simeq 5000$ with a strict $k_\mathrm{cut}\simeq 1.5\;{\rm Mpc^{-1}}$, while in the noBNT case a $k_\mathrm{cut}$ at the same scale with $\mathcal{T}_\mathrm{FD}=10^{-2}$ leads to a maximum observable $\ell\simeq 300$.

In the initial version of this work, we found that applying the BNT transform to $\xi_\pm(\theta)$ did not yield results as effective as applying it to $C_\ell$. This is because, in configuration space, Fourier modes are mixed in the construction of the estimator, due to the Legendre transform, which converts $C_\ell$ into $\xi_\pm(\theta)$. This can be seen in Figure 1 of \cite{{2021A&A...645A.104A}} where scale cuts exhibit strong oscillatory behavior across a wide range of $\ell$ values.
For this reason, we chose to focus exclusively on the angular power spectrum.

In Section \ref{sec:results}, we will perform MCMC analysis, and the comparison of posterior contours for noBNT and BNT will further highlight the effect of scale leakage on parameter constraints.

\section{Results}
\label{sec:results}

\subsection{Parameters forecast: general setup}
\label{subsec:IVA}

To quantify the ability of BNT to measure cosmological parameters with optimized scale cuts, we will conduct a series of MCMC forecasts assuming a True Background Cosmology (TBC) and performing parameter measurements using various 3-dimensional power spectrum models that differ from the TBC model. Specifically, we aim to address two key points:

\begin{itemize}
\item What is the advantage of BNT over noBNT in terms of information retention, for different choices of $k_{\rm cut}$ and $\mathcal{T}_\mathrm{FD}$?
\item To what extent can BNT still measure cosmological parameters with precision, without significant information loss, when the assumed power spectrum model deviates from the TBC?
\end{itemize}

These comparisons will be presented in section \ref{subsec:IVB}; here, we outline the general setup of the analysis. It is important to note that all BNT/noBNT comparisons are performed on a fair basis: nuisance parameters are marginalized consistently, and the selection of cosmological parameters remains the same. This approach is motivated by our aim to examine the $S_8$ tension in the context of BNT, though the applications of BNT extend far beyond this specific issue. For example, BNT could be employed to explore distinctive features of cosmological models, such as structures in the mass power spectrum as probes of the nature of dark matter.
In our analyses, we strive to remain agnostic about the true nonlinear information in our universe, focusing instead on determining whether the true $S_8$ value can be accurately recovered within a $1\sigma$ confidence interval.

\begin{table}[t]
\centering
\caption{\label{tbl:prior}%
The prior of all the parameters used in our \textsc{Nautilus} sampling. $\mathcal{U}(a,b)$ is a flat prior 
between $a$ and $b$, and $\mathcal{N}(c,d)$ is a Gaussian prior that centred at $c$ with standard deviation $d$. For cases where intrinsic alignments, multiplicative biases and/or redshift error parameters are not sampled, their values are set to the `Fiducial Value' column which indicates the values taken by the parameters for the fiducial cosmology.}

\rowcolors{2}{white}{lightgray!20}

\begin{tabular}{|l|c|c|}
\hline
\rowcolor{cyan!10} 
\ \textrm{Parameters} \ &
\ \textrm{Fiducial Value} \ & 
\textrm{Priors} \\
\hline
\ $\Omega_m$ & $0.2905$ & $\mathcal{U}(0.1,0.6)$\\
\ $10^9 A_s$ & $2.1868$ & $\mathcal{U}(1.5,5.0)$\\
\ $\Omega_b$ & $0.0473$ & $\mathcal{U}(0.03,0.07)$\\
\ $n_s$ & $0.9690$ & $\mathcal{U}(0.92,1.02)$\\
\ $h$ & $0.6898$ & $\mathcal{U}(0.55,0.85)$\\
\ $A_\mathrm{B}$ & $3.1300$ & $\mathcal{U}(1.0,6.0)$\\
\ $A_\mathrm{IA}$ & $0.0000$ & $\mathcal{U}(-6.0,6.0)$\\
\ $\eta_\mathrm{IA}$ & $0.0000$ & $\mathcal{U}(-5.0,5.0)$\\
\hline
\ $m_i$ & $0.0000$ & $\mathcal{N}(0.0,0.0005)$ \\
\ $\delta_{z,1}$ & $-0.026$ & $\mathcal{N}(-0.026,0.0019)$ \\
\ $\delta_{z,2}$ & $0.0227$ & $\mathcal{N}(0.0227,0.0020)$ \\
\ $\delta_{z,3}$ & $-0.026$ & $\mathcal{N}(-0.026,0.0019)$ \\
\ $\delta_{z,4}$ & $0.0126$ & $\mathcal{N}(0.0126,0.0020)$ \\
\ $\delta_{z,5}$ & $0.0193$ & $\mathcal{N}(0.0193,0.0020)$ \\
\ $\delta_{z,6}$ & $0.0083$ & $\mathcal{N}(0.0083,0.0020)$ \\
\ $\delta_{z,7}$ & $0.0382$ & $\mathcal{N}(0.0382,0.0020)$ \\
\ $\delta_{z,8}$ & $0.0027$ & $\mathcal{N}(0.0027,0.0020)$ \\
\ $\delta_{z,9}$ & $0.0340$ & $\mathcal{N}(0.0340,0.0021)$ \\
\ $\delta_{z,10}$ & $0.0495$ & $\mathcal{N}(0.0495,0.0021)$ \\
\hline
\ $\sigma_8$ & $0.8256$ & \ Derived Parameter \ \\
\ $S_8$ & $0.8124$ & \ Derived Parameter \ \\
\hline
\end{tabular}
\end{table}

\begin{figure*}
\centering
\includegraphics[width=0.85\textwidth, trim = 0.25cm 0.2cm 0.2cm 0.25cm, clip]{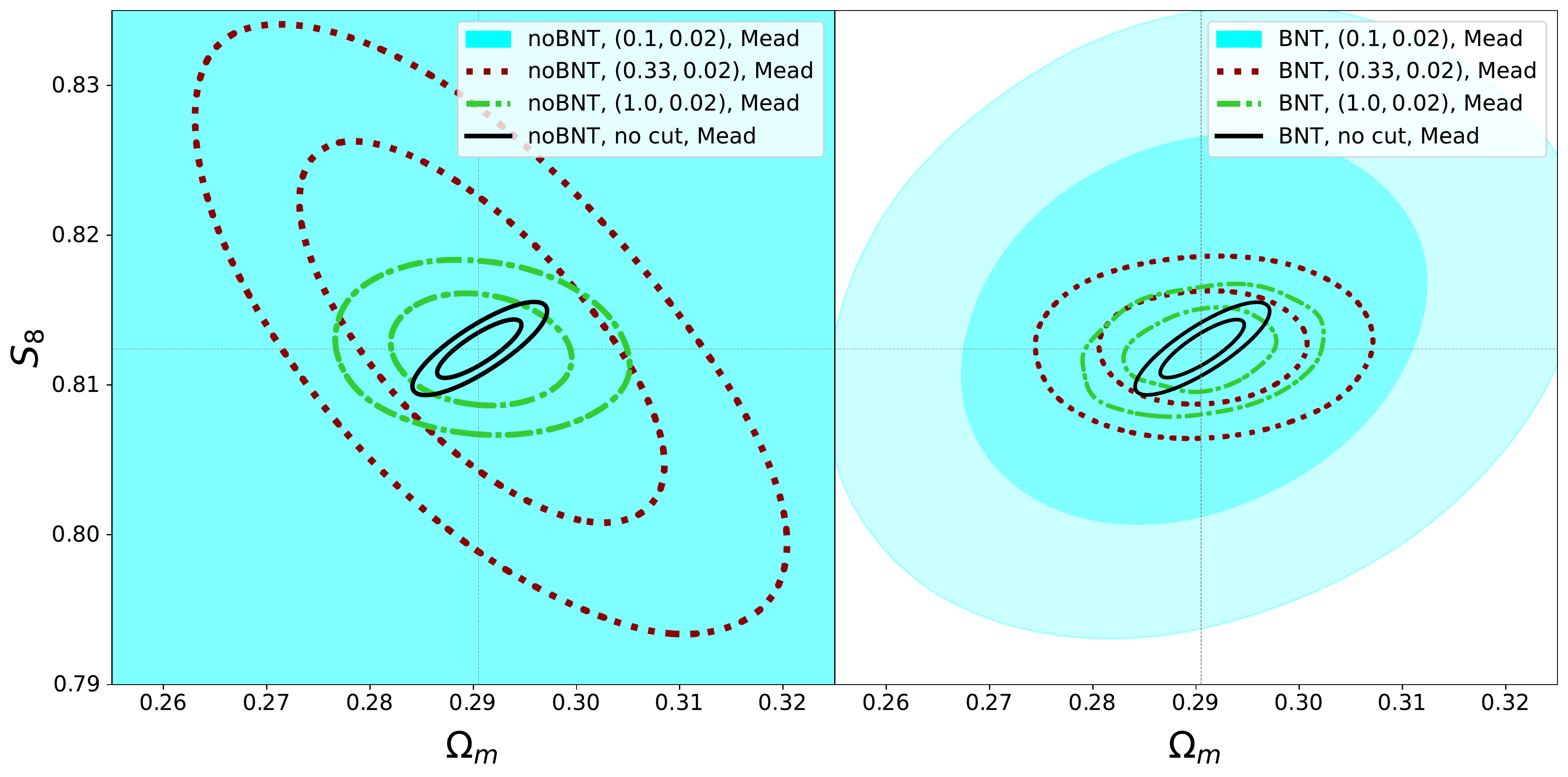}
\caption{Constraints in the $(\Omega_m,S_8)$ plane for the noBNT (left panel) and BNT (right panel). The power spectrum of all models are calculated using the \citet{HMcode16}-version of \textsc{HMcode}. Four posterior contours are represented: the {\color[HTML]{000000}\bf Black} solid line corresponds to the case where no $k_{\rm cut}$ and no $\mathcal{T}_\mathrm{FD}$ are applied. The {\color[HTML]{31CD31}\bf Lime Green}, {\color[HTML]{8B0000}\bf Dark Red} and {\color[HTML]{00AFFF}\bf Cyan} contours correspond to $k_{\rm cut}=(1.0,0.33,0.1)\;{\rm Mpc^{-1}}$ respectively, and fixed $\mathcal{T}_\mathrm{FD}=0.02$. Note that the {\color[HTML]{00AFFF}\bf Cyan} contours for noBNT and $k_{\rm cut}=0.1\;{\rm Mpc^{-1}}$ exceeds the plot boundary.}
\label{fig:kcut_comparison}
\end{figure*}

\begin{figure*}
\centering
\includegraphics[width=0.85\textwidth, trim = 0.25cm 0.2cm 0.2cm 0.25cm, clip]{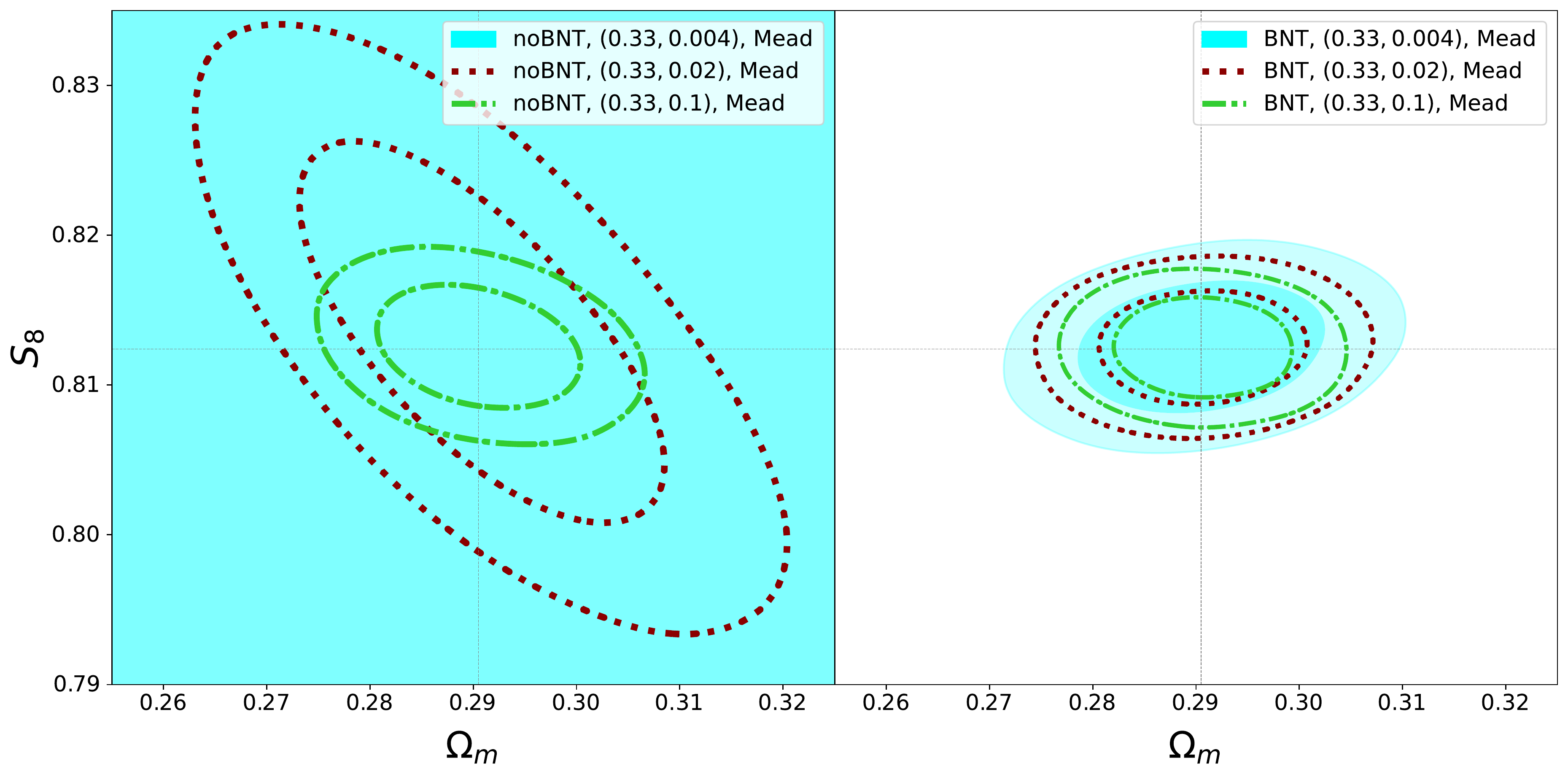}
\caption{Sam as Figure \ref{fig:kcut_comparison} with three posterior contours: the {\color[HTML]{31CD31}\bf Lime Green}, {\color[HTML]{8B0000}\bf Dark Red} and {\color[HTML]{00AFFF}\bf Cyan} contours correspond to $\mathcal{T}_\mathrm{FD}=(0.1,0.02,0.004)$ respectively, and fixed $k_{\rm cut}=0.33\;{\rm Mpc^{-1}}$. Note that the {\color[HTML]{00AFFF}\bf Cyan} contours for noBNT and $\mathcal{T}_\mathrm{FD}=0.004$ exceeds the plot boundary.}
\label{fig:TFD_comparison}
\end{figure*}

The TBC setup is given as follow:

\begin{itemize}
    \item We will utilize several common nonlinear models: \textsc{HMcode}-2016 \cite{HMcode16}, \textsc{Halofit}-2012 \cite{Takahashi12}, \textit{Baryon Correction Model} \cite{BCM1,BCM2}, and \textsc{AxionHMcode} \cite{AxionHMcode}. To generate Markov Chain cosmologies, we employed the standard \textsc{HMcode}-2016 as a baseline, running it against all the nonlinear models mentioned above.
    \item We use 50 $\ell$-bins logarithmically spaced from $\ell_{\rm min} = 50$ to $\ell_{\rm max} = 5000$, along with the redshift distribution shown in Figure \ref{fig:nz}. 
    \item The data vectors of the models used in the posterior sampling process are computed with the Core Cosmology Library (CCL) \cite{Pyccl}. CCL is a publicly available, standardized library for performing basic calculations of various cosmological observables, developed and calibrated by the LSST Dark Energy Science Collaboration (DESC) \cite{2018arXiv180901669T}.
    \item The covariance matrices are computed using \textsc{OneCovariance} \cite{OneCov}, a tool optimized for calculating covariance matrices from two-point statistics in photometric large-scale structure surveys. The statistical noise terms in the covariance matrix were calculated using a survey area $A_\mathrm{eff} = 15000\;\mathrm{deg}^2$, a number density of galaxies per tomographic bin $n_{\rm gal} = 3.0$, and the ellipticity noise $\sigma_e = 0.3$\cite{2020A&A...642A.191E}.
    \item We compute the parameter posteriors using the \textsc{Nautilus} \cite{Nautilus} package. \textsc{Nautilus} is a highly efficient importance-nested sampling toolkit for Bayesian posterior and evidence estimation, utilizing deep learning techniques. The sampling includes five cosmological parameters, one halo model parameter, two intrinsic alignment parameters, shear multiplicative bias and redshift error parameters for each tomographic redshift bin. Detailed prior information is provided in Table \ref{tbl:prior}. All these parameters are used in the conventional fiducial setup, both with and without the scale cut. For the cases where the fiducial cosmology differs from the sampling computations, both multiplicative biases and redshift errors are fixed to expedite the sampling process, we have verified that the results do not change significantly.
\end{itemize}

\subsection{Constraining power on $(S_8,\Omega_m)$}
\label{subsec:IVB}

We begin by focusing on the joint constraints on $(S_8, \Omega_m)$, comparing BNT and noBNT approaches for different choices of $(k_{\rm cut}, \mathcal{T}_\mathrm{FD})$.

Figure \ref{fig:kcut_comparison} presents the posterior of the two-dimensional parameter space $(S_8, \Omega_m)$, with $\mathcal{T}_\mathrm{FD}$ fixed at $0.02$ and three different $k_\mathrm{cut}$ values: $1.0$ Mpc$^{-1}$, $0.33$ Mpc$^{-1}$, and $0.1$ Mpc$^{-1}$. The left panel corresponds to the noBNT case, while the right panel shows the BNT case. The black solid contour in both panels is identical and represents the conventional lensing analysis (with no scale cuts and marginalization over nuisance parameters, as described in Section \ref{subsec:IVA}). It is evident that the contour size increases significantly for decreasing $k_{\rm cut}$ in the noBNT case, while the change is much smaller in the BNT case. In fact, for $k_{\rm cut} = 0.1\;{\rm Mpc^{-1}}$, the noBNT contours exceed the boundary of the plot. By comparison, the information loss relative to the no-cut case is much less significant for the BNT case. This demonstrates that if the mass power spectrum model is suspected to be biased in some $k$ range (e.g. for $k>0.1\;{\rm Mpc^{-1}}$), the BNT approach provides a more optimal method than the noBNT approach for removing small scales. It is often assumed that in the noBNT approach, marginalization over nuisance parameters can adequately address potential modeling biases in the mass power spectrum. However, as we will show in Section \ref{subsec:IVC}, this assumption is not valid. Not only is the information loss significant, but the resulting marginalized contours from the noBNT analysis remain biased. Overall, the degradation of the contours for decreasing $k_{\rm cut}$ is expected, and the reason for the more dramatic degradation for the noBNT case comes from the fact that the scale leakage is more pronounced in that case, as shown in Figure \ref{fig:color_BNT}.
\begin{table}[t]
\caption{\label{tbl:scale_cuts}%
The $S_8$ errorbars of different scale cut setups. Measurements are drawn directly from the package \textsc{Getdist} \cite{getdist}. We will continue to use the ($k_\mathrm{cut}$,
$\mathcal{T}_\mathrm{FD}$) notation for all contours and tables in the subsequent text. All posteriors above are sampled using the same theoretical calculations setup only except for the $\ell$-bin removal determined by Equation \ref{eqn:ellcut}. The `True Cosmology' here are all have $S_8 = 0.8124$. `Vary $m_i$ and $\delta_{z,i}$' means those two sets of parameters are sampled over the Gaussian prior given in \ref{tbl:prior}, and `fix $m_i$ and $\delta_{z,i}$' means those parameters are fixed to the fiducial value in our sampling processes.} 
\begin{tabular}{|l|c|c|}
\hline
\rowcolor{cyan!10}
\ ($k_{\rm cut}$,
$\mathcal{T}_\mathrm{FD}$) &
noBNT $\sigma_{_{S_8}}$ &
\ BNT $\sigma_{_{S_8}}$ \ \\ 
\hline
\hline
\rowcolor{lightgray!20}
\multicolumn{3}{|c|}{Marginalized over  $m_i$ and $\delta_{z,i}$}\\
\hline
\ no cut & $0.0009$ & $0.0009$ \\ 
\hline
\rowcolor{lightgray!20}
\ (1.0,0.1) & $0.0016$ & $0.0016$ \\ 
\ (1.0,0.02) & $0.0022$ & $0.0016$ \\ 
\rowcolor{lightgray!20}
\ (1.0,0.004) & $0.0039$ & $0.0017$ \\ 
\hline
\ (0.33,0.1) & $0.0024$ & $0.0020$ \\ 
\rowcolor{lightgray!20}
\ (0.33,0.02) & $0.0075$ & $0.0023$ \\ 
\ (0.33,0.004) \  & $0.0300$ & $0.0026$ \\ 
\hline
\rowcolor{lightgray!20}
\ (0.1,0.1) & $0.0150$ & $0.0063$ \\
\ (0.1,0.02) & \ No Constraints \ & $0.0077$ \\ 
\rowcolor{lightgray!20}
\ (0.1,0.004) & No Constraints  & $0.0093$ \\ 
\hline\hline
\rowcolor{lightgray!20}
\multicolumn{3}{|c|}{Fixed $m_i$ and $\delta_{z,i}$}\\
\hline
\ no cut & $0.0012$ & $0.0012$ \\ 
\hline
\rowcolor{lightgray!20}
\ (1.0,0.1) & $0.0015$ & $0.0015$ \\ 
\ (1.0,0.02) & $0.0021$ & $0.0015$ \\ 
\rowcolor{lightgray!20}
\ (1.0,0.004) & $0.0038$ & $0.0016$ \\
\hline
\ (0.33,0.1) & $0.0024$ & $0.0018$ \\ 
\rowcolor{lightgray!20}
\ (0.33,0.02) & $0.0074$ & $0.0021$ \\ 
\ (0.33,0.004) & $0.0298$ & $0.0024$ \\
\hline
\rowcolor{lightgray!20}
\ (0.1,0.1) & $0.0150$ & $0.0061$ \\ 
\ (0.1,0.02) & No Constraints & $0.0075$ \\ 
\rowcolor{lightgray!20}
\ (0.1,0.004) & No Constraints & $0.0090$ \\ 
\hline
\end{tabular}
\end{table}

\begin{figure*}[htbp]
\centering
\includegraphics[width=0.95\textwidth, trim = 0.25cm 0.2cm 0.2cm 0.25cm, clip]{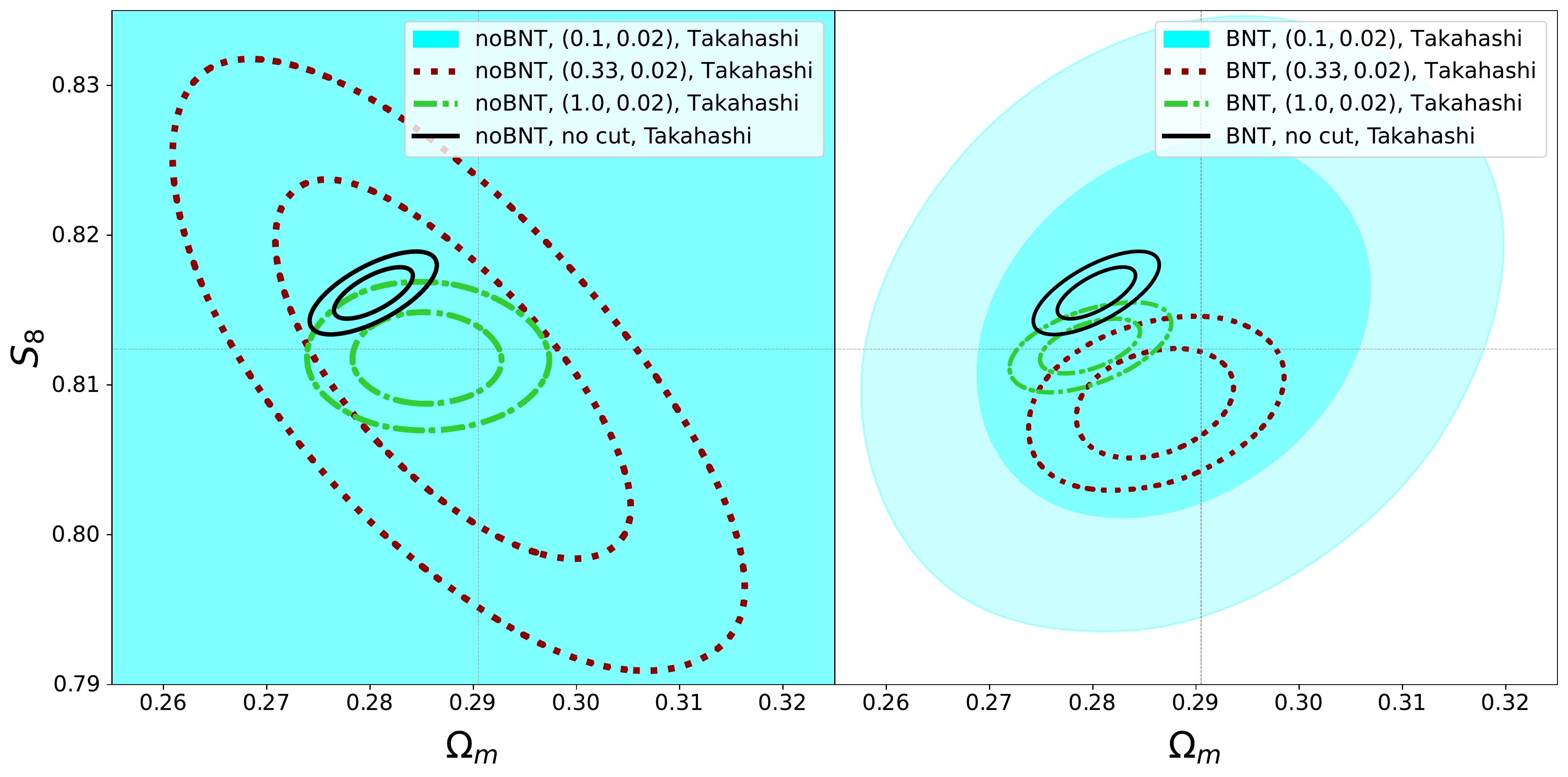}
\caption{Same as Figure \ref{fig:kcut_comparison}, except that the mock $C_\ell$ observations are calculated with 
with \citet{Takahashi12}-version of \textsc{Halofit} while the sampling process uses \citet{HMcode16}-version of \textsc{HMcode} as in Figure \ref{fig:kcut_comparison}.} \label{fig:Takahashi_comparison}
\end{figure*}

\begin{figure*}[htbp]
\centering
\includegraphics[width=0.95\textwidth, trim = 0.25cm 0.2cm 0.2cm 0.25cm, clip]{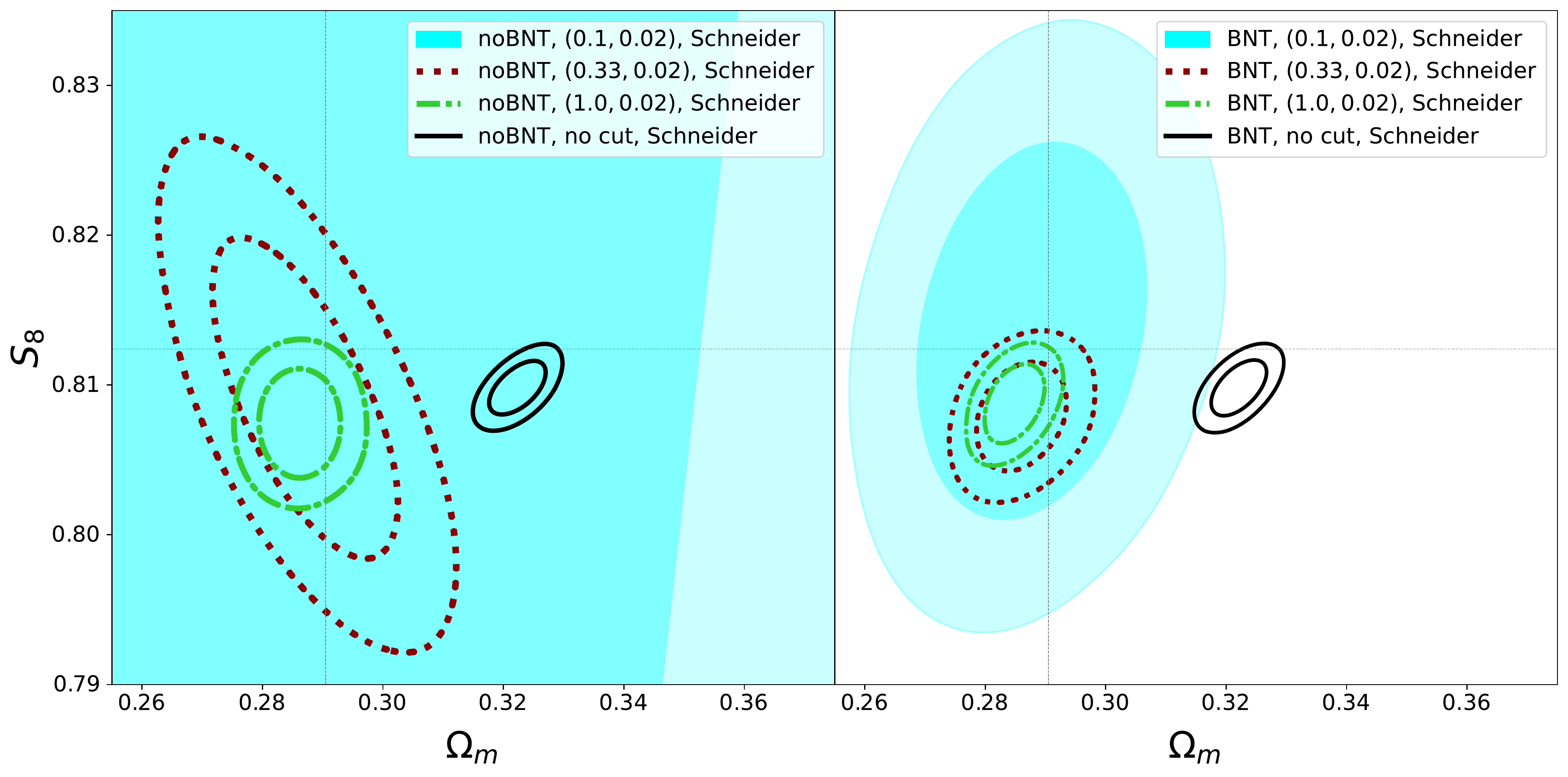}
\caption{Same as Figure \ref{fig:kcut_comparison}, except that the mock $C_\ell$ observations are calculated with 
with the Baryon Correction Model (\textsc{BCM}, \citet{BCM2}) while the sampling process uses \citet{HMcode16}-version of \textsc{HMcode} as in Figure \ref{fig:kcut_comparison}.
} 
\label{fig:Schneider_comparison}
\end{figure*}

\begin{figure*}[htbp]
\centering
\includegraphics[width=0.95\textwidth, trim = 0.25cm 0.2cm 0.2cm 0.25cm, clip]{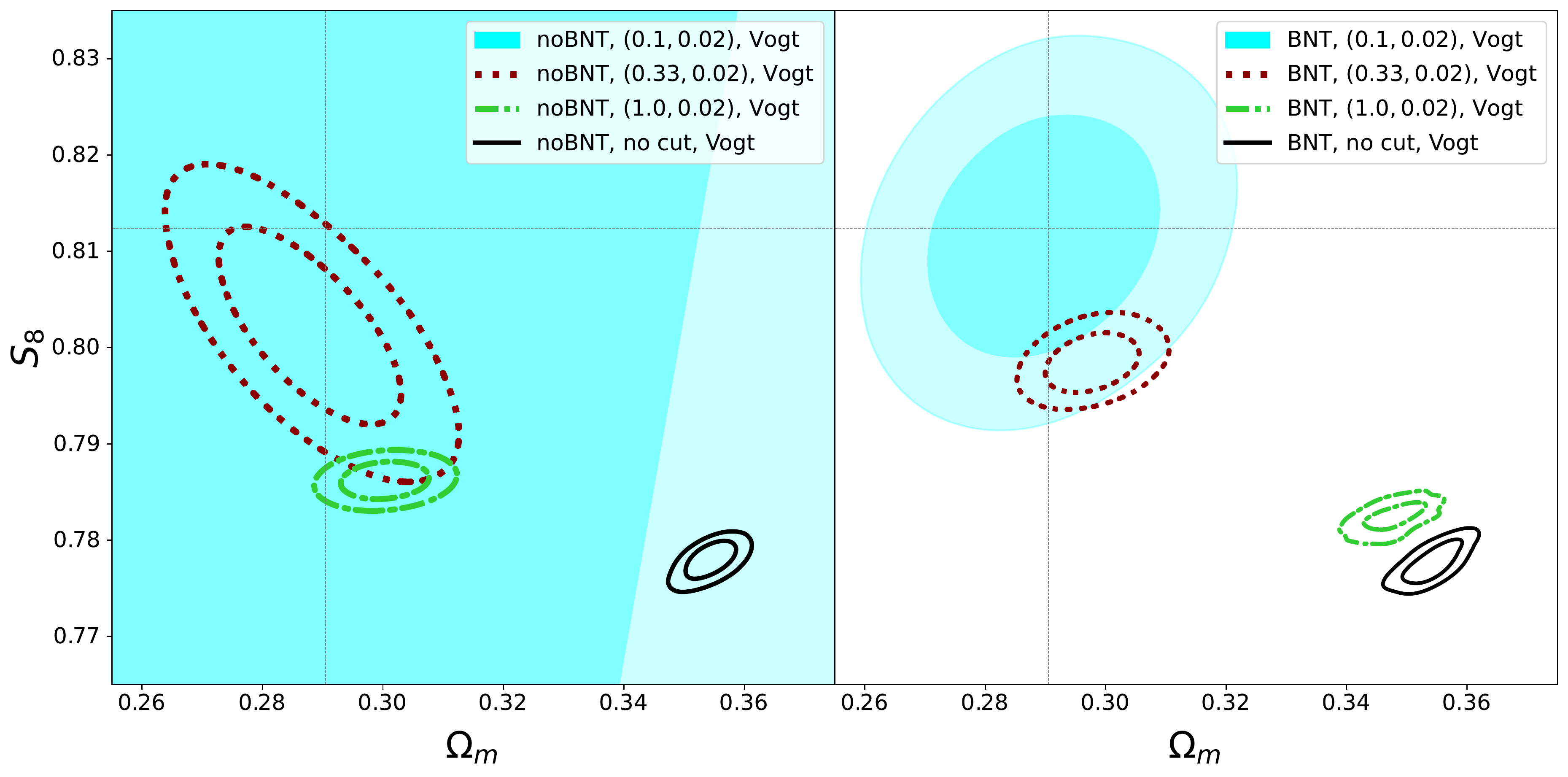}
\caption{Same as Figure \ref{fig:kcut_comparison}, except that the mock $C_\ell$ observations are calculated with 
with \textsc{AxionHMcode} (\citet{AxionHMcode}) while the sampling process uses \citet{HMcode16}-version of \textsc{HMcode} as in Figure \ref{fig:kcut_comparison}.}
\label{fig:Vogt_comparison}
\end{figure*}

Figure \ref{fig:TFD_comparison} presents the posterior of the two-dimensional parameter space $(S_8, \Omega_m)$, with $k_{\rm cut} = 0.33\;{\rm Mpc^{-1}}$ for three values of $\mathcal{T}_\mathrm{FD}$: $0.1$, $0.02$, and $0.004$. The increase in contour size for decreasing $\mathcal{T}_\mathrm{FD}$ is highly significant for the noBNT case, whereas it is surprisingly marginal for the BNT case. This behavior results from the sharp drop of $\hfR^{(a,b)} (\ell; k_{\rm cut}, \Pi^0)$ (Eq. \ref{Rfrak_def}), as illustrated in Figure \ref{fig:Cl}.
For the noBNT approach, minimising scale leakage requires both a decrease in $\mathcal{T}_\mathrm{FD}$ and an increase in $k_{\rm cut}$, as shown in Figure \ref{fig:color_BNT}. However, this comes at the cost of a substantial loss in constraining power within the $(S_8, \Omega_m)$ parameter space. In contrast, for the BNT approach, the scale leakage is almost independent of $\mathcal{T}_\mathrm{FD}$. As a result, the constraining power remains relatively unaffected, even with $k_{\rm cut} = 0.33\;\rm Mpc^{-1}$.

Clearly, the optimal choice of cuts $(k_{\rm cut}, \mathcal{T}_\mathrm{FD})$ depends on the accuracy of the power spectrum model. There is no universal rule that applies to all situations; in particular, the more reliable the model, the less necessary scale cuts should become. In Section \ref{subsec:IVC}, we will illustrate specific cases using different non-linear power spectrum models and an axion model where scale cuts might play a central role. Figures \ref{fig:kcut_comparison} and \ref{fig:TFD_comparison} demonstrate that, for a given $(k_{\rm cut}, \mathcal{T}_\mathrm{FD})$, the BNT approach consistently outperforms the noBNT approach. Table \ref{tbl:scale_cuts} summarizes the constraints on $S_8$ with both BNT and noBNT, for different nuisance parameter scenarios.

\subsection{Nonlinear Mass Power Spectrum}
\label{subsec:IVC}

In this section, we will derive the joint constraints on $(S_8,\Omega_m)$ when the mock observations are calculated with a different non-linear prescription for the power spectrum than the one used in the sampling of the posterior. We will investigate three cases: 1) \textsc{Halofit} \cite{Takahashi12} 2) Baryon Correction Model \cite{BCM2} and 3) \textsc{AxionHMcode} \cite{AxionHMcode}. The posterior sampling process still uses \textsc{HMcode} \cite{HMcode16}.

Figure \ref{fig:Takahashi_comparison} displays the same contours as Figure \ref{fig:kcut_comparison}, with the data vector calculated using the \citet{Takahashi12} version of \textsc{Halofit}. The solid black contours, representing the case without $k_{\rm cut}$ and $\mathcal{T}_\mathrm{FD}$ cuts, are identical for both the noBNT (left panel) and BNT (right panel) approaches. Although the bias in $S_8$ and $\Omega_m$ is small, it remains significant despite marginalization over $A_{\rm B}$.
Introducing $k_{\rm cut}$ reduces the bias as $k_{\rm cut}$ decreases, but the information loss becomes severe for $k_{\rm cut} \le 0.33\;{\rm Mpc^{-1}}$. For $(k_{\rm cut}, \mathcal{T}_\mathrm{FD}) = (0.1\;{\rm Mpc^{-1}}, 0.02)$, the BNT approach provides unbiased cosmological constraints, whereas the noBNT approach loses all constraining power.

Figure \ref{fig:Schneider_comparison} presents the $(S_8, \Omega_m)$ contours where the data vector is calculated using the Baryon Correction Model, designed to model a standard $\Lambda$CDM cosmology through profile-based calculations of various baryon feedback effects within dark matter halos. These effects include adiabatically relaxed dark matter, bound gas in hydrostatic equilibrium, gas ejected by supernovae and active galactic nuclei, and stellar components \cite{BCM2}. These feedback models introduce a cosmologically dependent suppression of the matter power spectrum relative to the dark-matter-only matter power spectrum, which is applied to the fiducial data vector in our analysis. The default BCM setups built in \textsc{Pyccl} was used, with the mass scale of hot gas suppression $M_c = 1.2\times10^{14}M_\odot$, ratio of escape to ejection radii $\eta_b = 0.5$, and the haracteristic scale (wavenumber) of the stellar component $k_s = 55\;$h Mpc$^{-1}$.
Figure \ref{fig:Schneider_comparison} exhibits a similar trend to Figure \ref{fig:Takahashi_comparison}, showing significant biases in the cosmological parameters, even after marginalization over nuisance parameters, when no cuts are applied. As $k_{\rm cut}$ decreases, the bias in cosmological parameters reduces, but the noBNT approach loses constraining power compared to the BNT approach.

Figures \ref{fig:Takahashi_comparison} and \ref{fig:Schneider_comparison} respectively address the theoretical uncertainty in the nonlinear modeling of the mass power spectrum and the physical uncertainty introduced by baryonic physics, as well as how these uncertainties can bias cosmological parameters. Another possibility is that dark matter itself affects the mass power spectrum, without direct observational evidence. This scenario arises in models such as fuzzy dark matter.
In Figure \ref{fig:Vogt_comparison}, we illustrate how using a fuzzy dark matter model for the data vector while employing conventional \textsc{HMcode} likelihood sampling can result in biased cosmological parameters. The fuzzy dark matter data vector is calculated with \textsc{AxionHMcode} \cite{AxionHMcode}, a nonlinear recipe calibrated to simulations involving ultralight axion-like particles mixed with cold dark matter. For this analysis, we use a setup with $m_\mathrm{axion} = 10^{-24}$ eV and $\Omega_\mathrm{axion} = 0.15\times \Omega_m = 0.0436$. This configuration represents a scenario where the role of dark matter in the nonlinear mass power spectrum is ignored, potentially contributing to the $S_8$ tension, as suggested by \cite{2022MNRAS.516.5355A} and \cite{2023MNRAS.525.4871G}.
The contours in Figure \ref{fig:Vogt_comparison} follow a similar pattern to those in Figures \ref{fig:Takahashi_comparison} and \ref{fig:Schneider_comparison}. The bias in $(S_8, \Omega_m)$ is significant but can be eliminated using the BNT approach, whereas the noBNT approach loses its constraining power. Notably, fuzzy dark matter leads to a lower amplitude of mass fluctuations, consistent with observations from stage III weak lensing surveys, which report a lower $S_8$ value compared to Planck results.

\begin{table*}[ht]
\rowcolors{2}{white}{lightgray!40}
\caption{\label{tbl:nl_recipes}%
Constraints on $S_8$, and posterior bias relative to the fiducial $S_8 = 0.8124$ and $\Omega_m = 0.2905$. The results are shown for the \textsc{Halofit}, BCM and \textsc{AxionHMcode} non-linear powerspectra, BNT and noBNT, and different values of $k_{\rm cut}$ and $\mathcal{T}_\mathrm{FD}$. In order to accelerate the calculations, the multiplicative bias and redshift errors are fixed to their fiducial values (see Table \ref{tbl:prior}). The resulting residual systematics never exceeds $10\%$ of the error on $S_8$. \textbf{\pos{Red}} Bias numbers denote the posterior medians that are higher than the fiducial $S_8$ and $\Omega_m$. \textbf{Black} Bias numbers means the posterior median is lower than those two true values.
All posteriors above are sampled using the same theoretical likelihood computation setup with only the $\ell_\mathrm{cut}$ changing, as determined by Equation \ref{eqn:ellcut}. 
For $(k_\mathrm{cut},\mathcal{T}_\mathrm{FD}) = (0.1,0.02)$ and $(0.1,0.004)$, all $\ell_\mathrm{cut}$ are below the minimum $\ell_{\rm min} = 50$ of the data vectors, therefore there are no noBNT constraints in those two cases.
}
\begin{tabular}{|>{\hspace{1mm}}l <{\hspace{1mm}}|>{\hspace{1mm}}c<{\hspace{1mm}} >{\hspace{1mm}}c<{\hspace{1mm}} | >{\hspace{1mm}}c<{\hspace{1mm}} >{\hspace{1mm}}c<{\hspace{1mm}} | >{\hspace{1mm}}c<{\hspace{1mm}} >{\hspace{1mm}}c<{\hspace{1mm}}|}
\hline
\rowcolor{cyan!20}
($k_\mathrm{cut}$,$\mathcal{T}_\mathrm{FD}$) & \textsc{Halofit} & Bias & BCM & Bias & \textsc{AxionHMcode} & Bias \\
\hline
no cut & $0.8162\pm0.0009$& $[\pos{4.2}\sigma_{S_8},5.1\sigma_{\Omega_m}]$ & $0.8098\pm0.0009$& $[2.9\sigma_{S_8},\pos{13.2}\sigma_{\Omega_m}]$ & $0.7776\pm0.0014$& $[24.9\sigma_{S_8},\pos{22.8}\sigma_{\Omega_m}]$\\
\hline
BNT, (1.0,0.1) & $0.8133\pm0.0010$& $[\pos{0.9}\sigma_{S_8},3.9\sigma_{\Omega_m}]$ & $0.8095\pm0.0015$& $[1.9\sigma_{S_8},1.7\sigma_{\Omega_m}]$ & $0.7798\pm0.0007$& $[48.3\sigma_{S_8},\pos{24.0}\sigma_{\Omega_m}]$\\ 
noBNT, (1.0,0.1) & $0.8118\pm0.0011$& $[0.5\sigma_{S_8},1.8\sigma_{\Omega_m}]$ & $0.8091\pm0.0015$& $[2.2\sigma_{S_8},1.2\sigma_{\Omega_m}]$ & $0.7806\pm0.0012$& $[26.5\sigma_{S_8},\pos{9.8}\sigma_{\Omega_m}]$\\ 
BNT, (1.0,0.02) & $0.8126\pm0.0011$& $[\pos{0.2}\sigma_{S_8},3.7\sigma_{\Omega_m}]$ & $0.8088\pm0.0015$& $[2.4\sigma_{S_8},1.9\sigma_{\Omega_m}]$ & $0.7825\pm0.0009$& $[32.5\sigma_{S_8},\pos{19.2}\sigma_{\Omega_m}]$\\ 
noBNT, (1.0,0.02) & $0.8118\pm0.0018$& $[0.3\sigma_{S_8},1.2\sigma_{\Omega_m}]$ & $0.8074\pm0.0021$& $[2.4\sigma_{S_8},1.0\sigma_{\Omega_m}]$ & $0.7862\pm0.0012$& $[21.8\sigma_{S_8},\pos{2.3}\sigma_{\Omega_m}]$\\ 
BNT, (1.0,0.004) & $0.8118\pm0.0012$& $[0.5\sigma_{S_8},3.5\sigma_{\Omega_m}]$ & $0.8086\pm0.0016$& $[2.4\sigma_{S_8},1.8\sigma_{\Omega_m}]$ & $0.7859\pm0.0012$& $[22.1\sigma_{S_8},\pos{14.0}\sigma_{\Omega_m}]$\\ 
noBNT, (1.0,0.004) & $0.8108\pm0.0044$& $[0.4\sigma_{S_8},0.5\sigma_{\Omega_m}]$ & $0.8083\pm0.0037$& $[1.1\sigma_{S_8},0.7\sigma_{\Omega_m}]$ & $0.7975\pm0.0035$& $[4.3\sigma_{S_8},0.5\sigma_{\Omega_m}]$\\ 
\hline
BNT, (0.3,0.1) & $0.8090\pm0.0018$& $[1.9\sigma_{S_8},1.2\sigma_{\Omega_m}]$ & $0.8077\pm0.0019$& $[2.5\sigma_{S_8},1.3\sigma_{\Omega_m}]$ & $0.7947\pm0.0014$& $[12.6\sigma_{S_8},\pos{1.7}\sigma_{\Omega_m}]$\\ 
noBNT, (0.3,0.1) & $0.8115\pm0.0024$& $[0.4\sigma_{S_8},0.9\sigma_{\Omega_m}]$ & $0.8077\pm0.0024$& $[2.0\sigma_{S_8},0.9\sigma_{\Omega_m}]$ & $0.7903\pm0.0016$& $[13.8\sigma_{S_8},\pos{0.8}\sigma_{\Omega_m}]$\\ 
BNT, (0.3,0.02) & $0.8088\pm0.0022$& $[1.6\sigma_{S_8},1.0\sigma_{\Omega_m}]$ & $0.8079\pm0.0021$& $[2.1\sigma_{S_8},1.0\sigma_{\Omega_m}]$ & $0.7985\pm0.0018$& $[7.7\sigma_{S_8},\pos{1.6}\sigma_{\Omega_m}]$\\ 
noBNT, (0.3,0.02) & $0.8109\pm0.0075$& $[0.2\sigma_{S_8},0.3\sigma_{\Omega_m}]$ & $0.8091\pm0.0064$& $[0.5\sigma_{S_8},0.4\sigma_{\Omega_m}]$ & $0.8022\pm0.0061$& $[1.7\sigma_{S_8},0.3\sigma_{\Omega_m}]$\\ 
BNT, (0.3,0.004) & $0.8090\pm0.0025$& $[1.4\sigma_{S_8},0.7\sigma_{\Omega_m}]$ & $0.8084\pm0.0025$& $[1.6\sigma_{S_8},0.7\sigma_{\Omega_m}]$ & $0.8007\pm0.0023$& $[5.1\sigma_{S_8},\pos{1.3}\sigma_{\Omega_m}]$\\ 
noBNT, (0.3,0.004) & $0.7974\pm0.0350$& $[0.4\sigma_{S_8},0.2\sigma_{\Omega_m}]$ & $0.8053\pm0.0293$& $[0.3\sigma_{S_8},\pos{0.0}\sigma_{\Omega_m}]$ & $0.8076\pm0.0247$& $[0.2\sigma_{S_8},\pos{0.1}\sigma_{\Omega_m}]$\\
\hline
BNT, (0.1,0.1) & $0.8130\pm0.0061$& $[\pos{0.1}\sigma_{S_8},0.4\sigma_{\Omega_m}]$ & $0.8129\pm0.0061$& $[\pos{0.1}\sigma_{S_8},0.4\sigma_{\Omega_m}]$ & $0.8110\pm0.0061$& $[0.2\sigma_{S_8},0.1\sigma_{\Omega_m}]$\\ 
noBNT, (0.1,0.1) & $0.8102\pm0.0150$& $[0.1\sigma_{S_8},\pos{0.0}\sigma_{\Omega_m}]$ & $0.8090\pm0.0150$& $[0.2\sigma_{S_8},\pos{0.0}\sigma_{\Omega_m}]$ & $0.8076\pm0.0150$& $[0.4\sigma_{S_8},\pos{0.0}\sigma_{\Omega_m}]$\\ 
BNT, (0.1,0.02) & $0.8139\pm0.0075$& $[\pos{0.2}\sigma_{S_8},0.2\sigma_{\Omega_m}]$ & $0.8138\pm0.0075$& $[\pos{0.2}\sigma_{S_8},0.2\sigma_{\Omega_m}]$ & $0.8118\pm0.0075$& $[0.1\sigma_{S_8},0.0\sigma_{\Omega_m}]$\\ 
BNT, (0.1,0.004) & $0.8145\pm0.0089$& $[\pos{0.2}\sigma_{S_8},0.1\sigma_{\Omega_m}]$ & $0.8142\pm0.0089$& $[\pos{0.2}\sigma_{S_8},0.1\sigma_{\Omega_m}]$ & $0.8123\pm0.0090$& $[0.0\sigma_{S_8},0.0\sigma_{\Omega_m}]$\\
\hline
\end{tabular}
\end{table*}

Table \ref{tbl:nl_recipes} summarizes all constraints on $S_8$ for the various $(k_\mathrm{cut}, \mathcal{T}\mathrm{FD})$ configurations used in this work and the three non-linear power spectrum models. The {\it Bias} columns indicate the systematic shifts from the fiducial cosmology for $(S_8, \Omega_m)$. All calculations presented in Table \ref{tbl:nl_recipes} were obtained by marginalizing over $A_{\mathrm{B}}$, a nonlinear nuisance parameter describing baryonic feedback in \textsc{HMcode}, as used in cosmic shear analyses conducted by KiDS, DES, and HSC. Instead of the conventional prior $A_\mathrm{B} \in (2.0, 3.13)$, we adopt a much wider prior of $A_\mathrm{B} \in (1.0, 6.0)$.
Despite this broader prior, the fiducial $S_8$ and $\Omega_m$ cosmology is not recovered within $5\sigma$ in any scenario where no scale cut is applied, regardless of the non-linear power spectrum recipe used. All our calculations indicate that marginalizing over nuisance parameters alone may not be an effective strategy for Stage-IV lensing surveys to mitigate nonlinear biases, as our understanding of the nonlinear universe remains incomplete.

\section{Discussion and Conclusions}

Scale cuts in weak lensing data, whether applied in configuration space or harmonic space, are a critical component of weak lensing studies aimed at addressing cosmological parameter tensions, such as $S_8$. However, a thorough characterization of the impact of theoretical biases on cosmological analysis and the reliability of the traditional scale cuts approach remains lacking for Stage IV surveys.

In this paper, we investigate how a linear transformation of the weak lensing data vector, referred to as the BNT transform, could offer an optimal solution for implementing scale cuts. We present a method based on the BNT transform applied in harmonic space, demonstrating that it is possible to define an $\ell$ cut that closely approximates a $k$ cut while effectively controlling high $k$ leakage. This level of precision is unattainable in the noBNT approach, which relies on the mean distance of the redshift bin to transform $k$ into $\ell$. Figure \ref{fig:Cl} illustrates our proposed method and provides a comparison between the BNT and noBNT approaches.

Our findings can be summarized as follows:

\begin{itemize}
    \item Theoretical biases in the mass distribution, particularly in the mass power spectrum $P(k)$ studied here, are unavoidable, leading to potential biases even at large scales, due to significant $k$ scale leakage. Since weak lensing observes mass in projection, these biases are distributed across all observed angular scales when traditional summary statistics are used (here angular scale refers to $\theta$ or $\ell$). We found that no angular scale is immune to these biases unless drastic $k$-space scale cuts are applied, which results in significant information loss. The corresponding scale leakage is illustrated in Figure \ref{fig:color_BNT}.
    
    \item Marginalizing over nuisance parameters, such as those modeling baryonic physics, may be insufficient to absorb residual biases and does not provide a diagnostic to detect biased results. We emphasize the importance of characterizing $k$-mode leakage as a diagnostic tool for identifying the location of potential biases in $k$-space.
    
    \item Using BNT-transformed data vectors, it is possible to apply an $\ell$-cut for each tomographic bin combination that effectively mimics a $k$-cut, referred to as $k_{\rm cut}$. High-$k$-mode leakage can be significantly reduced compared to traditional approaches by combining $\ell$-cuts with a threshold $\mathcal{T}_\mathrm{FD}$, which limits the fractional deviation from the uncut data vector. We found that BNT constraints are more robust against variations in $k_{\rm cut}$ and $\mathcal{T}_\mathrm{FD}$, while noBNT constraints degrade rapidly with decreasing $k_{\rm cut}$ and $\mathcal{T}_\mathrm{FD}$.
    
    \item In recent years, some studies have suggested that the $S_8$-tension might originate from nonlinear effects. In standard cosmic shear analyses, these nonlinear contributions are marginalized using nuisance parameters, such as the halo concentration parameter $A_\mathrm{B}$, which models baryon feedback in \textsc{HMcode}. Using three representative scenarios—\textsc{Halofit}, the Baryon Correction Model (BCM), and \textsc{AxionHMcode}—we mimic potential ignorance of the true nonlinear universe. Our results show that likelihood analyses without scale cuts fail to recover the $\Omega_m-S_8$ median within $3\sigma$, even after marginalization over $A_\mathrm{B}$ using a wide, uninformative prior. In contrast, analyses restricted to the linear regime with the BNT Transform successfully capture the $\Omega_m-S_8$ median across all three nonlinear scenarios, with error bars comparable to standard BNT likelihood results. This approach effectively halves the uncertainty relative to Stage-III or Planck outcomes.
\end{itemize}

These findings highlight the importance of employing the BNT methodology for Stage-IV lensing studies in order to mitigate nonlinear systematics and provide unbiased cosmological constraints. This paper serves as a proof of concept, and follow-up studies include:

\begin{itemize}
    \item A proper inclusion of intrinsic alignment is missing. The BNT implementation we have discussed uses groups of three tomographic bins to define the new lensing kernels, therefore, the contribution of intrinsic alignment should only be limited to these three tomographic bins. This means that IA will not mix the scales across all tomographic bins and our conclusions should not dramatically change.
    \item Adding CMB lensing as a high redshift source plane would be an interesting addition to the BNT transform. More theoretical development would be in order since CMB lensing is a scalar and a study of how to perform BNT with a mix of scalar and pseudo-vectors has not been done.
    \item We have shown how to use BNT to remove scales that could bias cosmological parameters. It is in principle possible to use BNT to optimize the analysis of $P(k)$ in some $k$ interval and target specific mass power spectrum features, e.g. fuzzy dark matter or Baryon Acoustic Oscillations \cite{2024PhRvL.132x1001F}
    \item BNT does not have to be restricted to the angular power spectrum $C_\ell$. It can be applied to higher-order statistics, mass maps, or any other lensing estimator built from first-order shear or convergence.
\end{itemize}

\begin{acknowledgments}
All theoretical calculations are based on \textsc{Pyccl}\cite{Pyccl} and we present our whole sampling code suit in \cite{BNT_repo}. 
We calculate errorbars and plot posterior contours plotting with \textsc{Getdist}\cite{getdist} package. We are grateful to Joachim Harnois-D\'eraps, Hendrik Hildebrandt, Mike Hudson, Benjamin Joachimi, Hiranya Peiris, Keir Rogers, Cora Uhlemann, Angus Wright, and the whole German Centre for Cosmological Lensing (GCCL) for their constructive feedbacks during the construction of the entire structure for the study and the analysis. We thank Sophie Vogt for her help with \textsc{AxionHMcode}. We thank Robert Reischke for providing his covariance matrix calculation code \textsc{OneCovariance} with technical supports. We are particularly grateful to S\'ebastien Fabbro for his help with the \textsc{canfar} scientific calculation service provided by the Canadian Astronomical Data Centre (CADC). RD acknowledges support from the Outer Space Institute through the MINDS Space Security Network grant.
\end{acknowledgments}

\appendix

\section{How to calculate $\delta\chi^2$}
\label{appendix:deltachi2}

In the appendix, we give a mathematical prescription of how to calculate the incremental $\chi^2$, or $\delta\chi^2$, as an $C_\ell$-based method to infer the $k$-mode dependence of the constraining power. 

The definition of the $\delta\chi^2$ term is written in Equation \ref{chi_incr}:
\be
\delta\chi^2_{[k_1,k_2]}(\Pi) \equiv \delta\chi^2_{[k_1,k_2],\mathrm{auto}}(\Pi) + \delta\chi^2_{[k_1,k_2],\mathrm{cross}}(\Pi).
\nonumber
\ee

where the auto-correlation term $\delta\chi^2_{[k_1,k_2],\mathrm{auto}}(\Pi)$ and cross-correlation term $\delta\chi^2_{[k_1,k_2],\mathrm{cross}}(\Pi)$ are defined in the Equation \ref{chi2_sum}:
\bea
\delta\chi^2_{[k_1,k_2],\mathrm{auto}}(\Pi)&=& \nonumber \\
\Big[
\Delta {\cal C}^{^T}_{[k_1,k_2]}(\Pi) &\cdot& \mathbb{C}^{-1}\cdot \Delta {\cal C}_{[k_1,k_2]}(\Pi)\Big] \\
\delta\chi^2_{[k_1,k_2],\mathrm{cross}}(\Pi)&=& \nonumber \\
\Big[
\Delta {\cal C}^{^T}_{[k_1,k_2]}(\Pi) &\cdot& \mathbb{C}^{-1} \cdot \sum_{\substack{[k'_1,k'_2]\ne \\  [k_1,k_2]}}\Delta {\cal C}_{[k'_1,k'_2]}(\Pi).
\Big]
\eea

The calculation in Eq.\ref{chi_incr} requires the estimation of $\Delta{\cal C}_{[k_1,k_2]}(\Pi)$ as defined in Eq.\ref{Delta_Cdef}. It is useful to introduce the weight function ${\cal W}_{[k_1,k_2]}(\Pi^0)$ and the quantity $\Delta{\cal C}^0_{[k_1,k_2]}(\Pi)$:

\bea
{\cal W}_{[k_1,k_2]}(\Pi^0)&\equiv& \dfrac{{\cal C}_{[k_1,k_2]}(\Pi^0)}{{\cal C}(\Pi^0)}\nonumber\\
\Delta{\cal C}^0_{[k_1,k_2]}(\Pi) &\equiv& {\cal W}_{[k_1,k_2]}(\Pi^0)\Delta{\cal D}(\Pi)
\label{DeltaC_approx}
\eea
where ${\cal C}(\Pi^0) = {\cal C}_{[k_\mathrm{min},k_\mathrm{max}]}(\Pi^0)$ represents the tomographic-binned vector covering the entire $k$-domain.
We adopt the approximation $\Delta{\cal C}_{[k_1,k_2]}(\Pi) \approx \Delta{\cal C}^0_{[k_1,k_2]}(\Pi)$, which simplifies the computational complexity from $O(m \times n)$ to $O(m + n)$.
This significantly improves efficiency, particularly when evaluating multiple cosmologies.
To validate this approximation, we performed several tests using random cosmologies and $k$-bins. The fractional difference in $\delta\chi^2$ values computed using $\Delta{\cal C}_{[k_1,k_2]}(\Pi)$ and $\Delta{\cal C}^0_{[k_1,k_2]}(\Pi)$ consistently remained within $\pm 5\%$, and in most cases, it was below $\pm 1\%$. The derivation and practical details of the $\delta\chi^2$ calculations are discussed in Appendix \ref{appendix:deltachi2}.

We can then further speed up the calculation with the approximation from Equation \ref{DeltaC_approx}:

\bea
\delta\chi^2_{[k_1,k_2],\mathrm{auto}}(\Pi) &=& \Delta {\cal C}^{^T}_{[k_1,k_2]}(\Pi) \cdot \mathbb{C}^{-1} \cdot \Delta {\cal C}_{[k_1,k_2]}(\Pi) \nonumber\\
&\simeq& 
\left[ {\cal W}_{[k_1,k_2]}(\Pi^0) \Delta{\cal D}(\Pi)\right]^{^T} \cdot \mathbb{C}^{-1} \cdot \nonumber\\
&&\left[ {\cal W}_{[k_1,k_2]}(\Pi^0) \Delta{\cal D}(\Pi)\right]\label{fb:dchi2auto}
\eea

The calculations for the auto-correlation term can also be applied with a $\ell_\mathrm{cut}$ truncation using the formalism described in Eq. \ref{star_eqn}. Under that case, the Eq. \ref{fb:dchi2auto} will be transformed to:
\bea
\delta\chi^2_{[k_1,k_2],\mathrm{auto}}(\Pi) &=& \left[ {\cal W}_{[k_1,k_2]}(\Pi^0) \Delta{\cal D^*}(\Pi)\right]^{^T} \cdot (\mathbb{C}^*)^{-1} \cdot \nonumber\\
&&\left[ {\cal W}_{[k_1,k_2]}(\Pi^0) \Delta{\cal D^*}(\Pi)\right]\label{fb:dchi2auto_ellcut}.
\eea

The same $\Delta{\cal D} \rightarrow \Delta{\cal D^*}$ and $\mathbb{C} \rightarrow \mathbb{C}^*$ transform also needs to be applied to the cross-correlation term calculation below if a $\ell_\mathrm{cut}$ is present. However, the cross-correlation term is not straightforward to be calculated directly with that time-saving approximation. Therefore, if we need to avoid using summation signs, we can then calculate a residual term:
\bea
\delta\chi^2_{[k_1,k_2],\mathrm{res}}(\Pi) &=& \chi^2(\Pi) -
\left[\Delta{\cal D}(\Pi)-\Delta{\cal C}_{[k_1,k_2]}(\Pi)\right]^{^T} \cdot \nonumber\\
&&\mathbb{C}^{-1} \cdot \left[\Delta{\cal D}(\Pi)-\Delta{\cal C}_{[k_1,k_2]}(\Pi)\right] \nonumber\\
&\simeq& \chi^2(\Pi) - \left([1-{\cal W}_{[k_1,k_2]}(\Pi^0)] \Delta{\cal D}(\Pi)\right)^{^T} \cdot \nonumber\\
&&\mathbb{C}^{-1} \cdot \left([1-{\cal W}_{[k_1,k_2]}(\Pi^0)] \Delta{\cal D}(\Pi)\right) \label{fb:dchi2res}.
\eea
  
From a geometric perspective, by visualising the $k$-bin correlations as a $k-k'$ two-dimensional space, one may be able to see this residual term as $\delta\chi^2_\mathrm{res} = \delta\chi^2_\mathrm{auto} + 2\delta\chi^2_\mathrm{cross}$. This intuition can also be proved mathematically:
\bea
\delta\chi^2_{[k_1,k_2],\mathrm{res}}(\Pi)
&=& \chi^2(\Pi) -\left[\Delta{\cal D}(\Pi)-\Delta{\cal C}_{[k_1,k_2]}(\Pi)\right]^{^T} \nonumber\\
&\cdot& \mathbb{C}^{-1} \cdot \left[\Delta{\cal D}(\Pi)-\Delta{\cal C}_{[k_1,k_2]}(\Pi)\right] \nonumber \\
&=& \chi^2(\Pi) - \chi^2(\Pi) - \delta\chi^2_{[k_1,k_2],\mathrm{auto}} \nonumber\\
&+& 2\left[\Delta{\cal C}^{^T}_{[k_1,k_2]}(\Pi) \cdot \mathbb{C}^{-1} \cdot \Delta{\cal D}(\Pi)\right]  \nonumber\\
&=& -\delta\chi^2_{[k_1,k_2],\mathrm{auto}} + 2\mathfrak{P}_1 \nonumber\\
&=& -\delta\chi^2_{[k_1,k_2],\mathrm{auto}} + 2\delta\chi^2_{[k_1,k_2],\mathrm{auto}} + 2\mathfrak{P}_2 \nonumber\\
&=& \delta\chi^2_{[k_1,k_2],\mathrm{auto}} + 2\delta\chi^2_{[k_1,k_2],\mathrm{cross}}.
\eea
where $\mathfrak{P}_{1}$ and $\mathfrak{P}_{2}$ are simplified notation for two inner products:
\bea
\mathfrak{P}_1 &=& \left[\Delta{\cal C}^{^T}_{[k_1,k_2]}(\Pi) \cdot \mathbb{C}^{-1} \cdot \sum_{[k_1',k_2']}\Delta {\cal C}_{[k_1',k_2']}(\Pi)\right] \nonumber\\
\mathfrak{P}_{2} &=& \left[\Delta{\cal C}^{^T}_{[k_1,k_2]}(\Pi) \cdot \mathbb{C}^{-1} \cdot \sum_{\substack{[k_1',k_2']\neq \\ [k_1,k_2]}}\Delta {\cal C}_{[k_1',k_2']}(\Pi)\right]
\eea
Therefore, in the real application, we calculate the incremental $\delta\chi^2$ by:
\be
\delta\chi^2_{[k_1,k_2]} = \frac{1}{2}\delta\chi^2_{[k_1,k_2],\mathrm{auto}} + \frac{1}{2}\delta\chi^2_{[k_1,k_2],\mathrm{res}}. \label{eqn:dchi2}
\ee

\bibliography{BNT_prd}

\end{document}